\documentclass[10pt,twocolumn,letterpaper]{article}

\usepackage{cvpr}              

\usepackage[most]{tcolorbox}

\newtcolorbox{disclaimerbox}{
    colback=gray!10,     
    colframe=gray!40,    
    boxrule=0.5pt,       
    arc=4pt,             
    auto outer arc,
    boxsep=5pt,
    left=6pt,
    right=6pt,
    top=4pt,
    bottom=4pt,
    enhanced jigsaw
}

\usepackage{lipsum}
\usepackage{comment}
\usepackage{svg}
\usepackage{tikz}
\usepackage{colortbl}
\usetikzlibrary{positioning, arrows.meta, chains, calc}

\definecolor{cvprblue}{rgb}{0.21,0.49,0.74}
\definecolor{lblue}{rgb}{0.9,0.95,1}
\definecolor{lpurple}{rgb}{0.35,0.25,0.55}
\definecolor{lgreen}{rgb}{0.95,1,0.95}
\definecolor{sblue}{rgb}{0,0.45,1}
\definecolor{lgray}{gray}{0.95}
\definecolor{lyellow}{rgb}{1,1,0.92}

\usepackage[pagebackref,breaklinks,colorlinks,citecolor=cvprblue]{hyperref}

\def\paperID{*****} 
\def\confName{CVPR}
\def\confYear{2025}

\title{NTIRE 2025 Challenge on RAW Image Restoration and Super-Resolution}

\author{
Marcos V. Conde~$^{*\dagger}$ \and
Radu Timofte~$^{*}$ \and
Zihao Lu~$^{*}$ \and
Xiangyu Kong\and
Xiaoxia Xing\and
Fan Wang\and
Suejin Han\and
MinKyu Park\and
Tianyu Zhang\and
Xin Luo\and
Yeda Chen\and
Dong Liu\and
Li Pang\and
Yuhang Yang\and
Hongzhong Wang\and
Xiangyong Cao\and
Ruixuan Jiang\and
Senyan Xu\and
Siyuan Jiang\and
Xueyang Fu\and
Zheng-Jun Zha\and
Tianyu Hao\and
Yuhong He\and
Ruoqi Li\and
Yueqi Yang\and
Xiang Yu\and
Guanlan Hong\and
Minmin Yi\and
Yuanjia Chen\and
Liwen Zhang\and
Zijie Jin\and
Cheng Li\and
Lian Liu\and
Wei Song\and
Heng Sun\and
Yubo Wang\and
Jinghua Wang\and
Jiajie Lu\and
Watchara Ruangsang\and
}

\begin{document}

\maketitle

\let\thefootnote\relax\footnotetext{$*$ Marcos V. Conde ($\dagger$ corresponding author, project lead), Zihao Lu, and Radu Timofte are the challenge organizers, while the other authors participated in the challenge and survey. \\
$*$~University of W\"urzburg, CAIDAS \& IFI, Computer Vision Lab.\\ 
NTIRE 2025 webpage:~\url{https://cvlai.net/ntire/2025}.\\
Code:~\url{https://github.com/mv-lab/AISP}} 

\begin{abstract}
This paper reviews the NTIRE 2025 RAW Image Restoration and Super-Resolution Challenge, highlighting the proposed solutions and results. New methods for RAW Restoration and Super-Resolution could be essential in modern Image Signal Processing (ISP) pipelines, however, this problem is not as explored as in the RGB domain. The goal of this challenge is two fold, (i) restore RAW images with blur and noise degradations, (ii) upscale RAW Bayer images by 2x, considering unknown noise and blur. In the challenge, a total of 230 participants registered, and 45 submitted results during thee challenge period. This report presents the current state-of-the-art in RAW Restoration.
\end{abstract}

\setlength{\abovedisplayskip}{1pt}
\setlength{\belowdisplayskip}{1pt}

\section{Introduction}

RAW Image Restoration has become an active topic in the computational photography field for enhancing image quality. This work focuses on three related sub-tasks: super-resolution, denoising, and deblurring. Each of these addresses image quality bottlenecks in portable camera devices: insufficient pixel density, excessive sensor noise, and motion blur due to long exposure times.

Although modern portable devices now feature image sensors with higher bit depths approaching those of professional-grade cameras, they remain constrained by physical size, power consumption, and thermal limitations. As a result, larger sensors and higher-quality optical components are impossible to be deployed in a limited space. Smaller sensor areas lead to reduced light collection per pixel and lower signal-to-noise ratios (SNR). Optical resolution and distortion resistance are also limited, making it difficult to achieve both high resolution and low noise simultaneously with great sharpness. These inherent limitations underline the irreplaceable role of image restoration tasks in portable imaging systems.

At the data level, RAW images are processed with only minimal operation, primarily linear amplification and white balance which preserving a nearly linear response to the sensor signal. In contrast, images processed through the ISP (Image Signal Processing) pipeline experience a series of strongly nonlinear operations such as demosaicing, tone mapping, gamma correction, and color adjustment \cite{xu2019rawsr, conde2024bsraw, conde2023perceptual, ignatov2021learnednpu, ignatov2020replacing}. These operations introduce irreversible information loss and amplify quantization errors and noise textures~\cite{brooks2019unprocessing, karaimer2016software}, constraining the performance for tasks like denoising and super-resolution when conducted in the sRGB domain. Although some recent works have demonstrated great performance on RAW image reconstruction from sRGB domain \cite{nam2017modelling, xing2021invertible}, reconstructed RAW image still have distance from the original RAW images, which means the reconstruction model couldn't restore the original detail pattern from the sRGB image, original RAW images still have more detail information for the model to learn.

Furthermore, due to the lack of a ISP-tuning standard and varying stylistic preferences across camera manufacturers, sRGB-format images from the same sensor can differ significantly more than their RAW images. This increases the difficulty in achieving generalization and robustness for cross-sensor model when relying on sRGB inputs.

At the same time, portable devices also suffer from limited computational resources. Therefore, model size and computational complexity must be prioritized factor for model design.
\vspace{2mm}

Thus, in this work we are presenting the solutions submitted for the NTIRE 2025 RAW Image Restoration and Super-Resolution Challenge. We are providing information regarding the challenge setup, with the task description and the challenge data properties characterizing the challenge dataset splits.  We are also listing information regarding the challenge participants, with their teams and affiliations. 

In \cref{sec:challenge} and \cref{sec:RAWIR_Challenge} we describe the challenge tracks, datasets and evaluations. In \cref{sec:srteams} and \cref{sec:irteams} we provide detailed descriptions of the best solutions.

\begin{figure*}[t]
    \centering
    \setlength{\tabcolsep}{1pt}
    \begin{tabular}{c c c c}
         \includegraphics[width=0.245\linewidth]{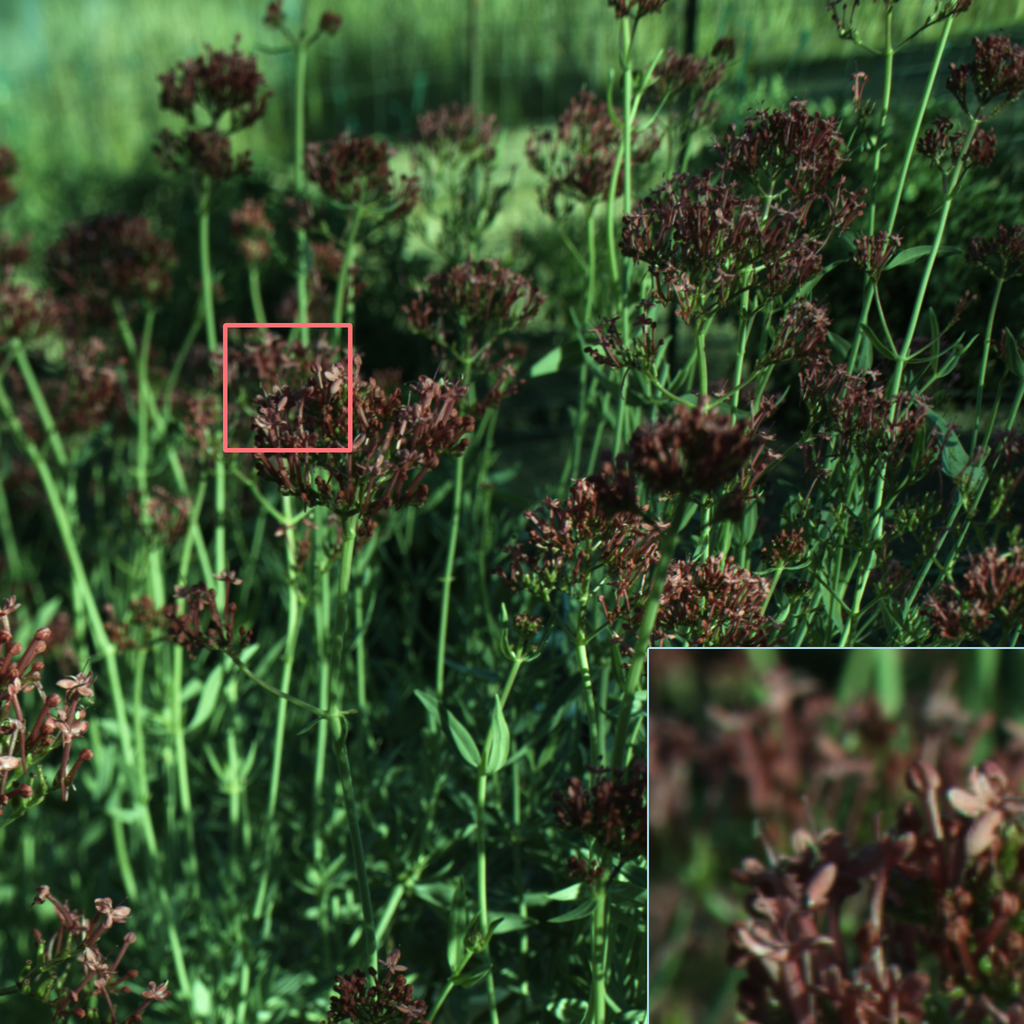} & 
         \includegraphics[width=0.245\linewidth]{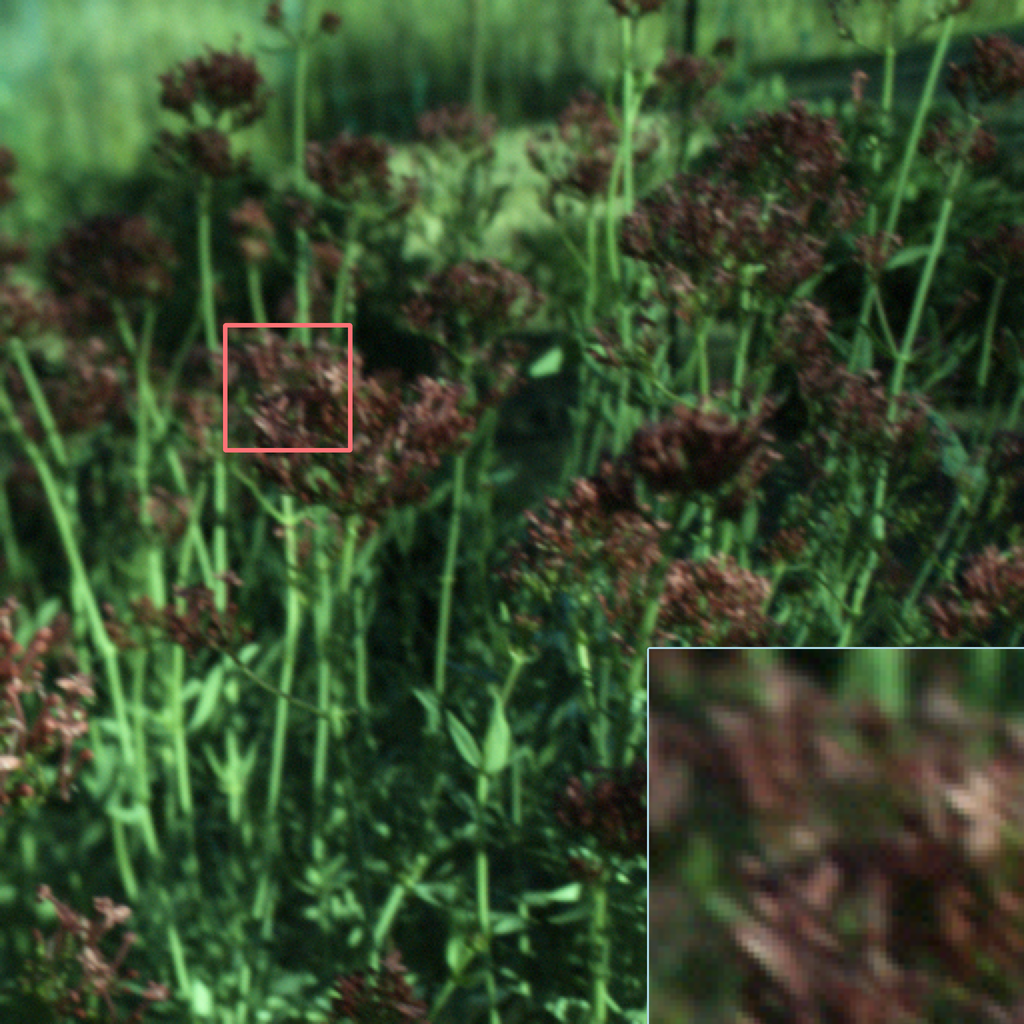} & 
         \includegraphics[width=0.245\linewidth]{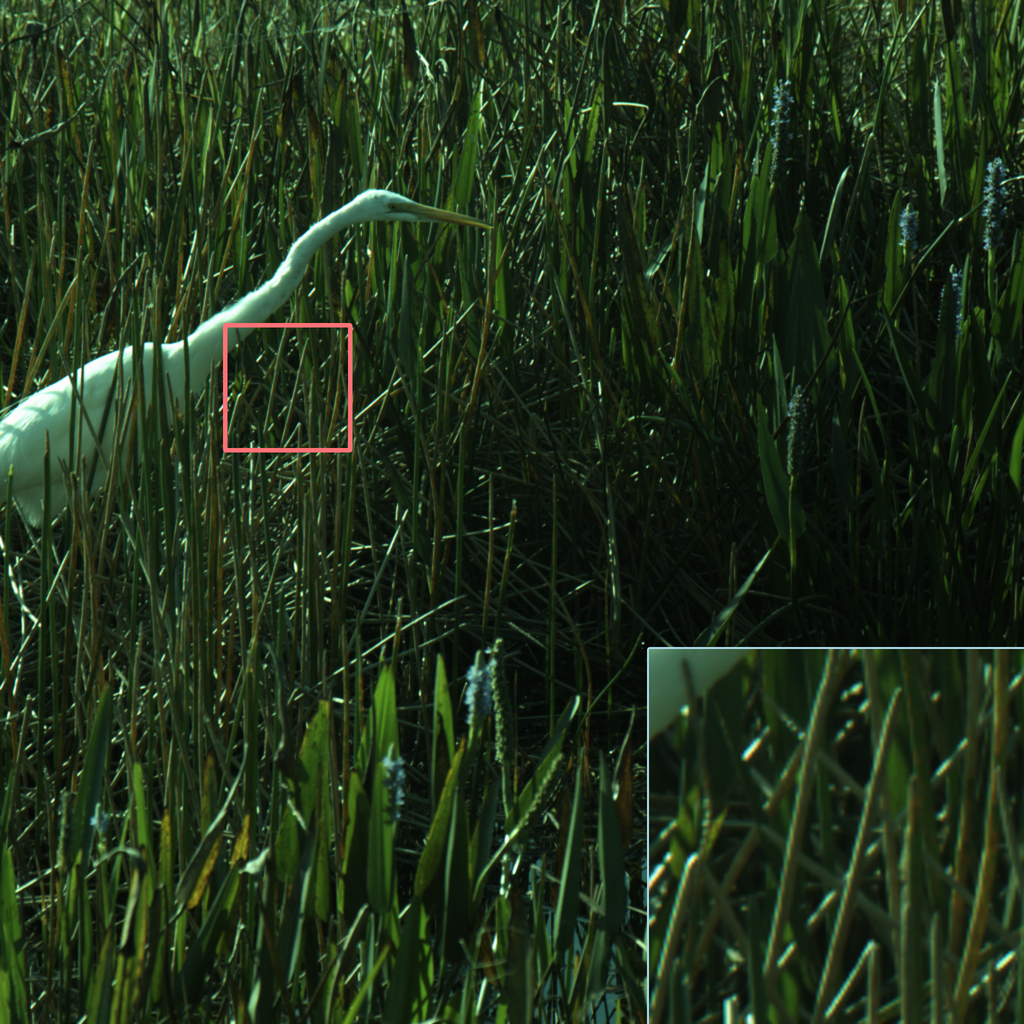} & 
         \includegraphics[width=0.245\linewidth]{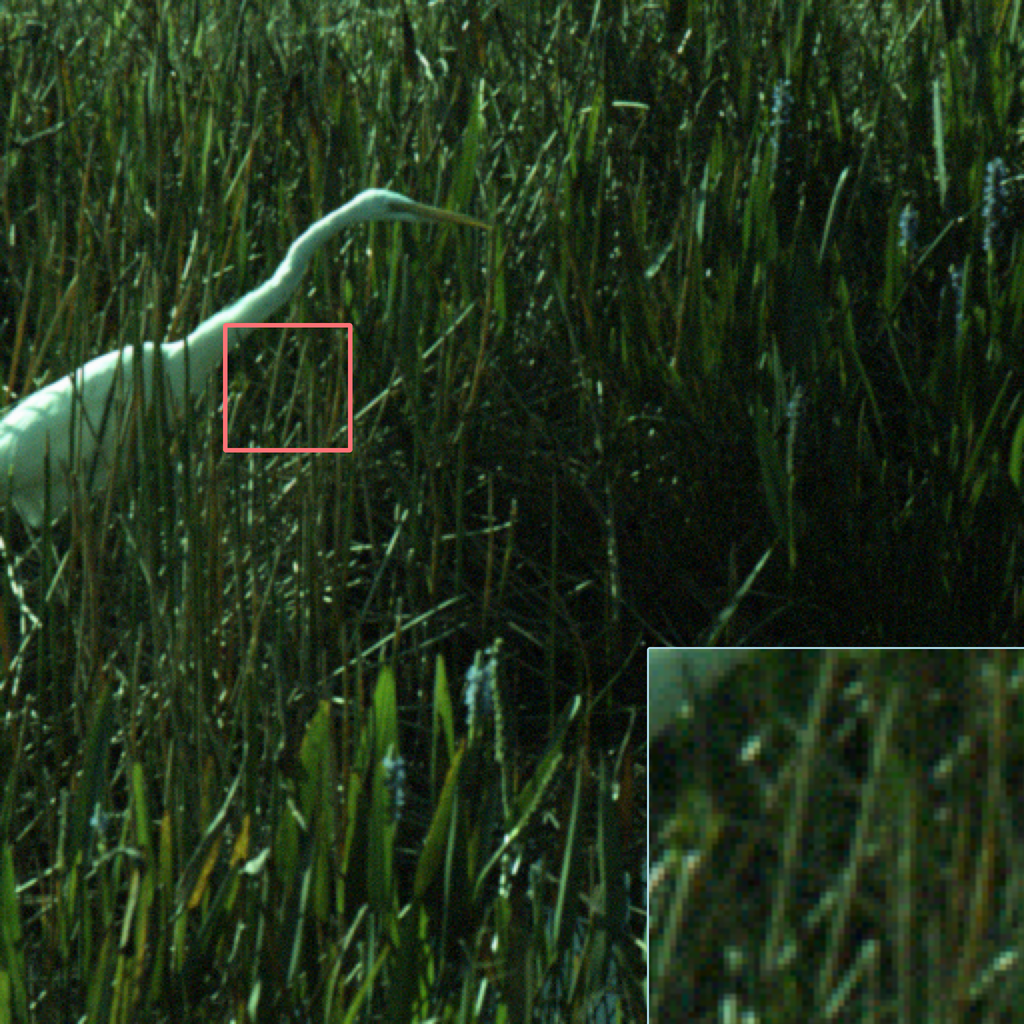}
         \tabularnewline
         \includegraphics[width=0.245\linewidth]{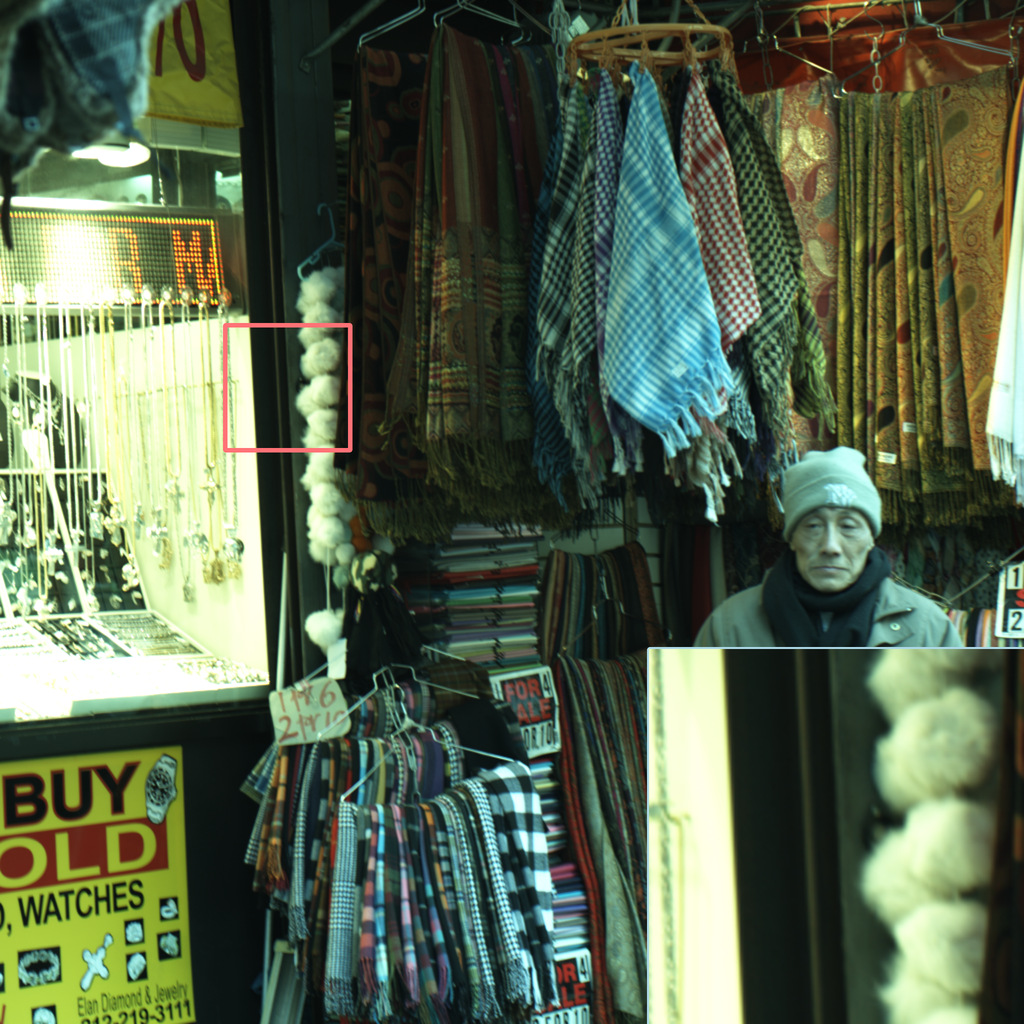} & 
         \includegraphics[width=0.245\linewidth]{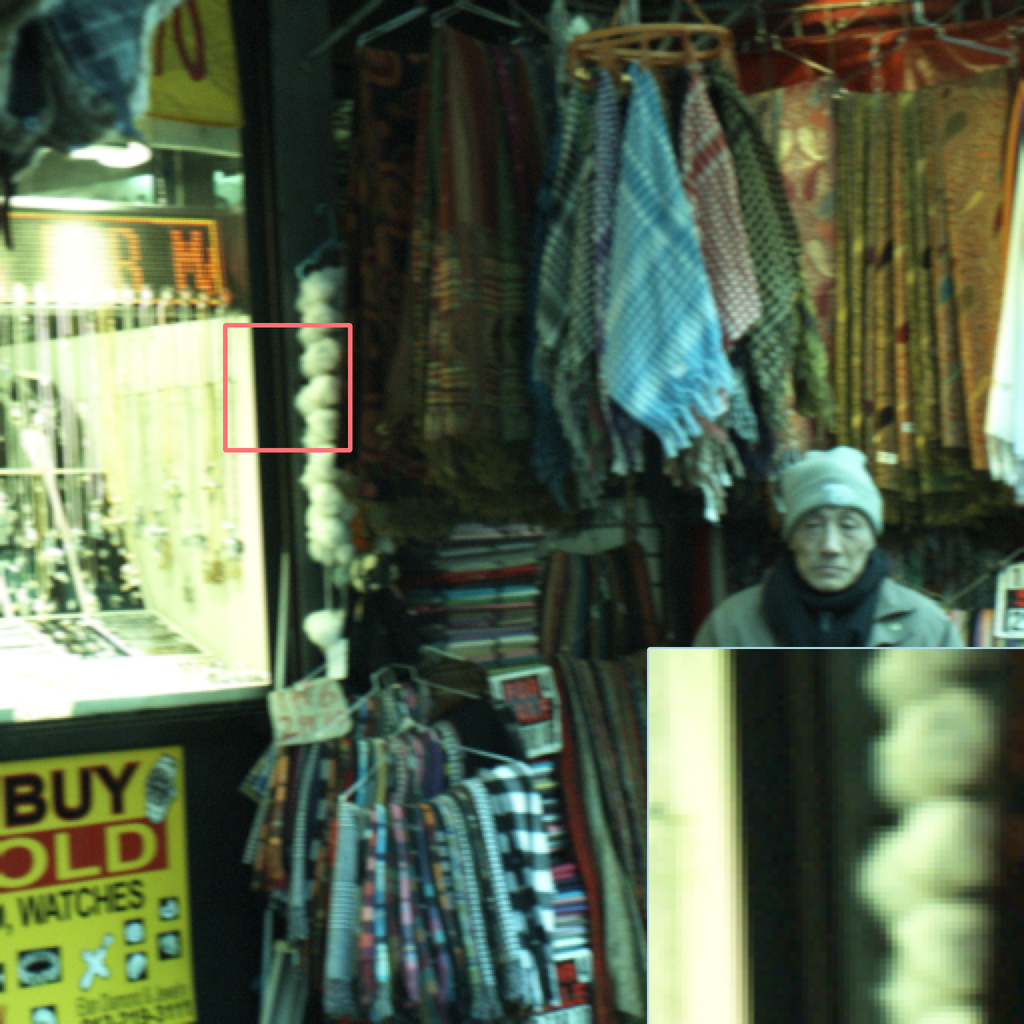} & 
         \includegraphics[width=0.245\linewidth]{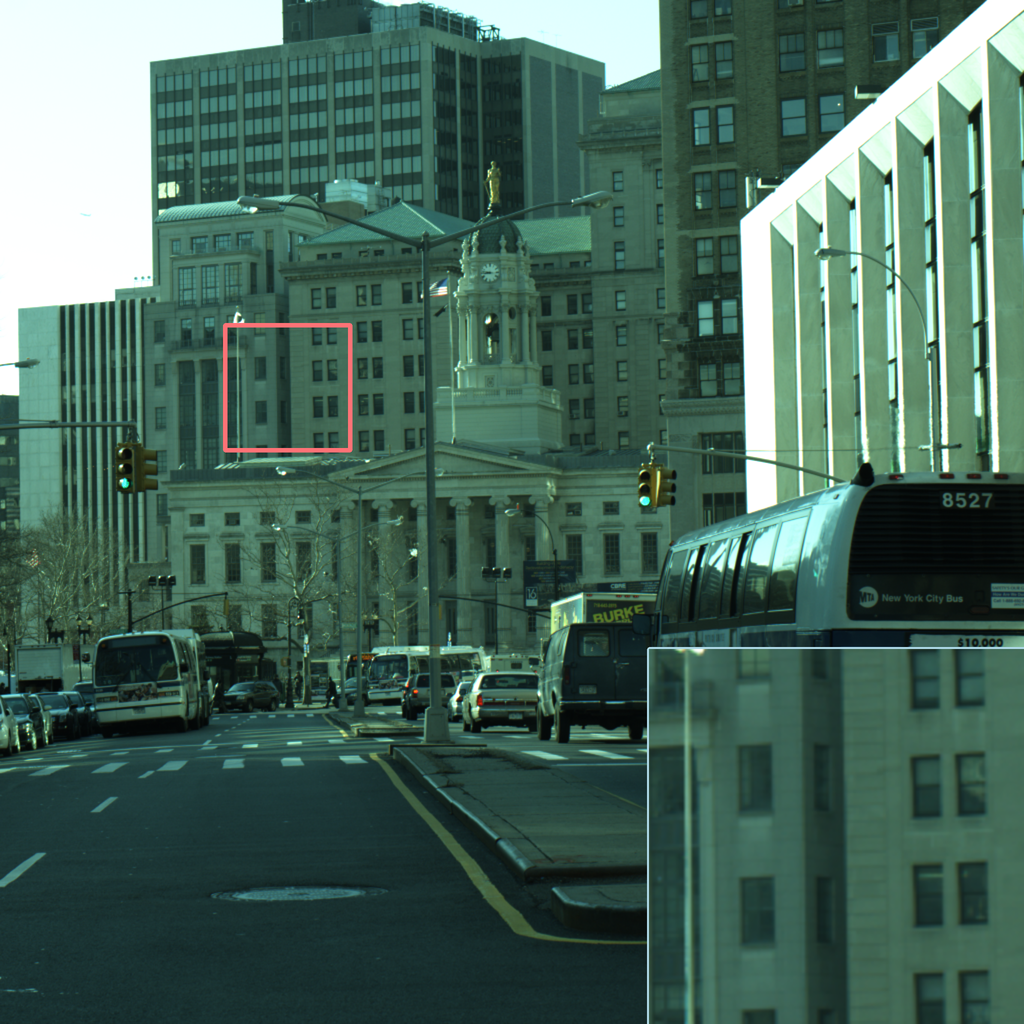} & 
         \includegraphics[width=0.245\linewidth]{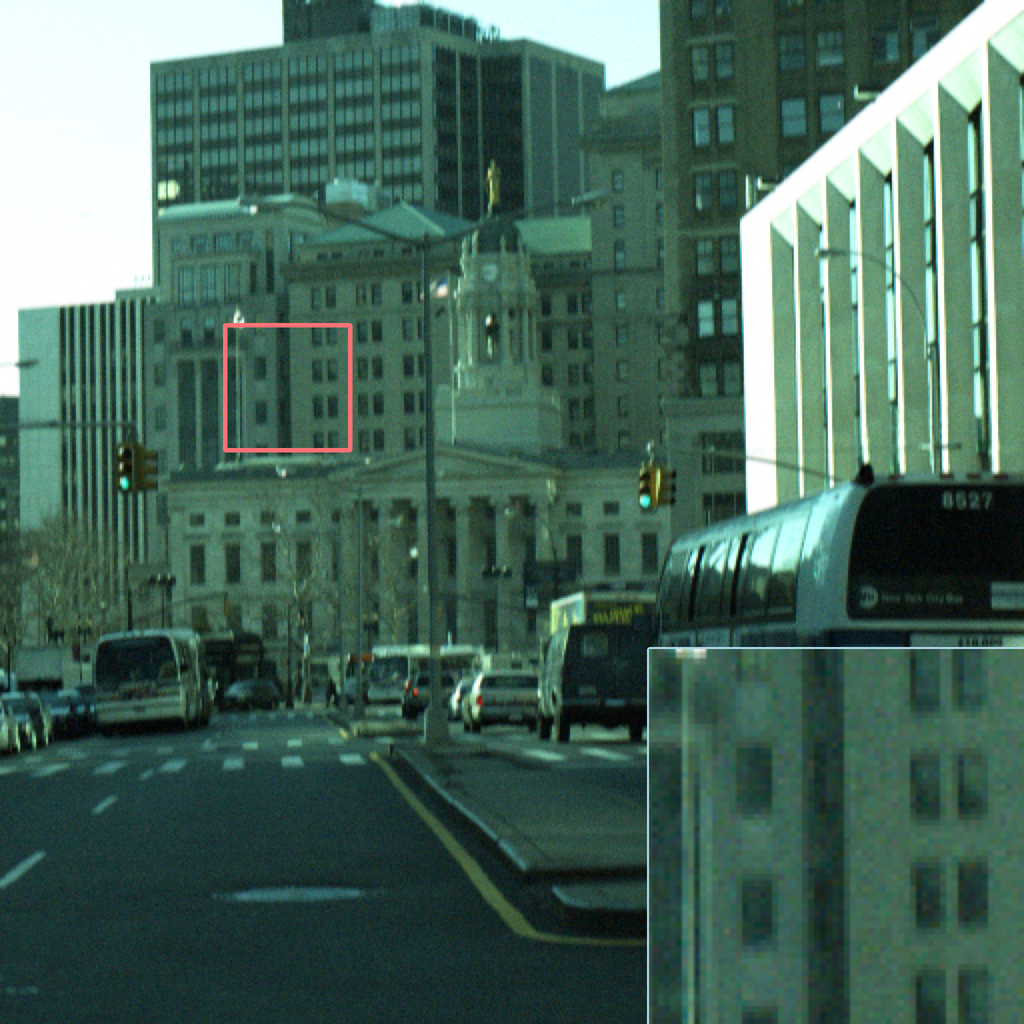}
         \tabularnewline
         HR Ground-truth & LR Input & HR Ground-truth & LR Input 
    \end{tabular}
    \caption{Samples of the \textbf{NTIRE 2025 RAW Image Super-Resolution Challenge} testing set.}
    \label{fig:test-samples}
    \end{figure*}


\paragraph{Related Computer Vision Challenges}

This challenge is one of the NTIRE 2025 \footnote{\url{https://www.cvlai.net/ntire/2025/}} Workshop associated challenges on: 
ambient lighting normalization~\cite{ntire2025ambient}, 
reflection removal in the wild~\cite{ntire2025reflection}, 
shadow removal~\cite{ntire2025shadow}, 
event-based image deblurring~\cite{ntire2025event}, 
image denoising~\cite{ntire2025denoising}, 
XGC quality assessment~\cite{ntire2025xgc}, 
UGC video enhancement~\cite{ntire2025ugc}, 
night photography rendering~\cite{ntire2025night}, 
image super-resolution (x4)~\cite{ntire2025srx4},
real-world face restoration~\cite{ntire2025face}, 
efficient super-resolution~\cite{ntire2025esr}, 
HR depth estimation~\cite{ntire2025hrdepth}, 
efficient burst HDR and restoration~\cite{ntire2025ebhdr}, 
cross-domain few-shot object detection~\cite{ntire2025cross}, 
short-form UGC video quality assessment and enhancement~\cite{ntire2025shortugc,ntire2025shortugc_data}, 
text to image generation model quality assessment~\cite{ntire2025text}, 
day and night raindrop removal for dual-focused images~\cite{ntire2025day}, 
video quality assessment for video conferencing~\cite{ntire2025vqe}, 
low light image enhancement~\cite{ntire2025lowlight}, 
light field super-resolution~\cite{ntire2025lightfield}, 
restore any image model (RAIM) in the wild~\cite{ntire2025raim}, 
raw restoration and super-resolution~\cite{ntire2025raw}, 
and raw reconstruction from RGB on smartphones~\cite{ntire2025rawrgb}.

\section*{Acknowledgments}
This work was partially supported by the Humboldt Foundation. We thank the NTIRE 2025 sponsors: ByteDance, Meituan, Kuaishou, and University of Wurzburg (Computer Vision Lab).

\section{NTIRE 2025 RAWSR Challenge}
\label{sec:challenge}

\begin{table}[!ht]
    \centering
    \resizebox{\linewidth}{!}{
    \begin{tabular}{r c c c c}
    \toprule
    \rowcolor{lgray} Method & Track & PSNR & SSIM & \#~Par. \\
    \midrule
    \rowcolor{lyellow} SMFFRaw-S~\ref{sec:xjtusr} & Efficient & 42.12 & 0.9433 & 0.18 \\
    RawRTSR~\ref{samsungsr}     & Efficient & 41.74 & 0.9417 & 0.19 \\
    NAFBN~\ref{sec:nju}            & Efficient & 40.67 & 0.9347 & 0.19 \\
    MambaIRv2~\ref{sec:tysl}  & Efficient & 40.32 & 0.9396 & 0.19 \\
    RepRAW-SR-Tiny~\ref{sec:eifflow}    & Efficient & 40.01 & 0.9297 & 0.02 \\
    RepRAW-SR-Large~\ref{sec:eifflow}    & Efficient   & 40.56 & 0.9339 & 0.09 \\
    ECAN~\ref{sec:cuee}      & Efficient & 39.13 & 0.9057 & 0.09 \\
    
    \midrule
    \rowcolor{lyellow} USTC~\ref{sec:ustc}    & General & 42.70 & 0.9479 & 1.94 \\     
    SMFFRaw-S~\ref{sec:xjtusr} & General   & 42.60 & 0.9467 & 1.99 \\
    RawRTSR-L~\ref{samsungsr}     & General   & 42.58 & 0.9475 & 0.26 \\
    ERBSFormer~\ref{sec:egroup}       & General & 42.45 & 0.9448 & 3.30 \\
    ER-NAFNet~\ref{ir_ER-NAFNet}     & General   & 41.17 & 0.9348 & - \\
    \midrule
    RBSFormer~\cite{jiang2024rbsformer} & 2024 & 43.649 & 0.987 & 3.3 \\
    BSRAW~\cite{conde2024bsraw} & 2024 & 42.853 & 0.986 & 1.5 \\
    Bicubic~\cite{conde2024ntire_raw} & 2024 & 36.038 & 0.952 & - \\
    \bottomrule
    \end{tabular}
    }
    
    \caption{\textbf{NTIRE 2025 RAWSR Results}. We provide \textbf{PSNR/SSIM} results on the complete testing set (200 images)~\cite{conde2024ntire_raw}. Efficient methods are constrained to maximum 200K parameters. As a reference, we provide the baselines from the NTIRE 2024 RAWSR Challenge~\cite{conde2024ntire_raw}. All the fidelity metrics are calculated in the RAW domain. We also report parameters (Par.) in millions. 
    }
    \label{tab:benchmark}
\end{table}

\subsection{Dataset}

The challenge dataset is based on BSRAW~\cite{conde2024bsraw} and the NTIRE 2024 RAWSR Challenge~\cite{conde2024ntire_raw}. Following previous work~\cite{ xu2019rawsr, xu2020exploiting, conde2024bsraw}, we use images from the Adobe MIT5K dataset~\cite{fivek}, which includes images from multiple Canon and Nikon DSLR cameras. 

The images are manually filtered to ensure diversity and natural properties (\ie remove extremely dark or overexposed images), we also remove the blurry images (\ie we only consider in-focus sharp images with low ISO).

The \textbf{pre-processing} is as follows: (i) we normalize all RAW images depending on their black level and bit-depth. (ii) we convert (``pack") the images into the well-known RGGB Bayer pattern (4-channels), which allows to apply the transformations and degradations without damaging the original color pattern information~\cite{liu2019learningrawaug}.

\noindent \textbf{Training:} We provide the participants 1064 images of resolution $1024 \times 1024 \times 4$, these are clean high-resolution (HR) RAW images. The LR degraded images can be generated on-line during training using the degradation pipeline proposed in BSRAW~\cite{conde2024bsraw}. 

Such degradation pipeline considers different noise profiles, multiple blur kernels (PSFs) and a simple downsampling strategy to synthesize low-resolution (LR) RAW images~\cite{conde2024rawir}. The participants can apply other augmentation techniques or expand the degradation pipeline to generate more realistic training data.

\subsection{Baselines}

We use BSRAW~\cite{conde2024bsraw} and previous years top-performing methods as references. The top performing challenge solutions improve the baseline performance, however, the neural networks are notably more complex in terms of design and computation.

\subsection{Results}

We use three testing splits: (i) Validation, 40 1024px images using during the model development phase. (ii) Test 1MP, 200 images of 1024px resolution. (iii) The same 200 test images at full-resolution $\approx12$MP. The participants process the corresponding LR RAW images (\eg $512 \times 512 \times 4$), and submit their results. Thus, the participants do not have access to the ground-truth images. 

We provide samples of the testing set in \cref{fig:test-samples}.

In \cref{tab:benchmark} we provide the challenge benchmark. Besides fidelity metrics such as PSNR and SSIM, we also provide relevant implementation details of each method. The methods can greatly improve the RAW images quality and resolution, even in the case of full-resolution 12MP images as output. 
All the proposed methods are able to increase the resolution and details of the RAW images while reducing blurriness and noise. Moreover, there are not detectable color artifacts. We can conclude that (synthetic) RAW image super-resolution can be solved similarity to RAW denoising. However, more realistic downsampling remains an open challenge.

\begin{table*}[ht]
    \centering
    \resizebox{\linewidth}{!}{
    \begin{tabular}{r  c  c c c  c c}
        \toprule
        \rowcolor{lgray} Method & Type & Test Level 1 & Test Level 2 & Test Level 3 & \# Params. (M) & \# MACs (G) \\
        \toprule
        Test Images   &   & 0.953 / 39.56 & 0.931 / 35.30 & 0.907 / 33.03 & & \\
        PMRID~\cite{}         & Baseline  & 0.982 / 42.41 & 0.965 / 38.43 & 0.951 / 35.97 & 1.032 & 1.21 \\
        NAFNET~\cite{chen2022simple}        & Baseline  & 0.983 / 43.50 & 0.972 / 39.70 & 0.962 / 37.49 & 1.130 & 3.99 \\
        MOFA~\cite{chen2023mofa}          & Baseline  & 0.982 / 42.54 & 0.966 / 38.71 & 0.974 / 36.33 & 0.971 & 1.14 \\
        \rowcolor{lgray} RawIR~\cite{conde2024rawir}  & Baseline  & 0.984 / 44.20 & 0.978 / 40.30 & 0.974 / 38.30 & 1.5 & 12.3 \\
        \midrule
        \midrule
        Samsung AI~\ref{sec:samsungir} & Efficient & \textbf{0.991 / 45.10} & \textbf{0.980 / 40.82} & \textbf{0.971 / 38.46} & 0.19 & 10.98 \\
        LMPR-Net~\ref{LMPR-Net}         & Efficient & 0.989 / 42.57 & 0.973 / 39.17 & 0.961 / 37.26 & 0.19 & 2.63 \\
        \midrule
        \midrule
        Samsung AI~\ref{sec:samsungir} & General & 0.993 / \textbf{46.04} & \textbf{0.985 / 42.25} & \textbf{0.978 / 40.10} & 4.97 & 23.79 \\
        Miers~\ref{sec:xiaomi-ir}           & General & \textbf{0.993} / 45.72 & 0.983 / 41.73 & 0.974 / 39.50 & 4.76 & \#N/A \\
        Multi-PromptIR~\ref{Multi-PromptIR}         & General & 0.986 / 44.80 & 0.978 / 41.38 & 0.968 / 38.96 & 39.92 & 158.24 \\
        ER-NAFNet~\ref{ir_ER-NAFNet}      & General & 0.992 / 45.10 & 0.972 / 39.32 & 0.953 / 36.13 & 4.57 & \#N/A \\
        \bottomrule
        \end{tabular}
        }
    \caption{\textbf{NTIRE 2025 RAW Image Restoration Results.} We provide the \textbf{SSIM/PSNR} results on the testing set (140 images) at different degradation levels. All the metrics are calculated in the RAW domain. We also provide the efficiency information about the models. All the best performance results are highlighted in bold.}
    \label{tab:rawir-benchmark}
\end{table*}

\section{NTIRE 2025 RAW Image Restoration (RAWIR) Challenge}
\label{sec:RAWIR_Challenge}

\subsection{Dataset}
\label{sec:IRDataset}

The RAW Image Restoration (RAWIR) challenge dataset uses images from diverse sensors including: Samsung Galaxy S9 and iPhone X (RAW2RAW~\cite{afifi2021semi}). Additionally, we collected images from Google Pixel 10, Vivo X90, and Samsung S21 to enrich the diversity of camera sensors and build a complete testing set. The dataset covers various scenes (indoor and outdoor), lighting conditions (day and night), and subjects. All the images were manually filtered to ensure high quality, sharpness and clear details \ie the images were captured under low ISO ($\leq$ 400), in-focus (without notable defocus or motion blur) and with proper exposure. The original RAW files were saved in DNG format, unprocessed by the smartphones' ISPs.

We applied the following \textbf{pre-processing} pipeline to the RAW images:
\begin{itemize}
    \item All images were normalized based on the camera black-level and bit depth (\eg 10, 12, 14 bits per pixel).
    \item The images were converted to the RGGB Bayer pattern (4-channels). 
    \item We crop the images into non-overlapping (packed) patches of dimension $512 \times 512 \times 4$.
\end{itemize}

We provide 2139 clean patches for \textbf{training} models. The participants use the baseline degradation pipeline~\cite{conde2024bsraw} to simulate realistic degradations. As part of the challenge, participants develop their own degradation pipelines to simulate more realistic blur and noise. The core components in the degradation pipeline includes different noise profiles, and multiple blur kernels (PSFs).

\subsection{Baselines}
\label{sec:IRBaseline}
We use PMRID~\cite{wang2020practical} and MOFA~\cite{chen2023mofa} as the main baseline model. 
We also include NAFNet~\cite{chen2022simple} into our baseline as a representative of UNet-like models. Both methods are popular efficient image denoising methods.

\subsection{Challenge Results}
\label{sec:IRResult}
The synthetic test dataset is generated by applying our degradation pipeline at different levels. The three degradation level settings are as follows: (i) $y=x+n$, the degradation is only sampled noise from real noise profiles. (ii) $y=(x*k)+n$, the degradation is noise and/or blur, with 0.3 probability of blur and 0.5 of real noise. (iii) $y=(x*k)+n$, all the images have realistic blur and noise.

In the \cref{tab:rawir-benchmark}, we provide the challenge benchmark and quantitative evaluation results of the challenge participants. The challenge covers a wide range of solutions: efficient (under 200K parameters), and general solutions without computational constraints. The team Samsung AI Camera wins the general and efficient track.

In \cref{fig:results2}, we provide qualitative results of the challenge methods. For the level 1 test set, most solutions showcase great performance by increasing the sharpness and reducing noise completely. For the levels 2 and 3, the solutions from team Samsung AI and Miers show the best results, while the output from the WIRTeam and ChickentRun show less effective blur removal. In general, since denoising and deblurring might require opposite operations, the models struggle to tackle both at the same time, specially blur.  

\vspace{2mm}

\begin{disclaimerbox}
In Section~\ref{sec:srteams} we describe the top solutions for \textbf{RAW Super-Resolution}. 

In the Section~\ref{sec:irteams} we describe the best solutions for \textbf{RAW restoration}. 

\vspace{2mm}

Note that the method descriptions were provided by each team as their contribution to this report.
\end{disclaimerbox}

\newpage
\section{RAW Image Super-Resolution Methods}
\label{sec:srteams}

\subsection{RawRTSR: Raw Real-Time Super Resolution}
\label{samsungsr}

\begin{center}

\vspace{2mm}
\noindent\emph{\textbf{Team Samsung AI}}
\vspace{2mm}

\author{Xiaoxia Xing~$^1$,
Fan Wang~$^1$,
Suejin Han~$^2$,
MinKyu Park~$^2$ \\
$^1$ Samsung R\&D Institute China - Beijing (SRC-B)\\
$^2$ The Department of Camera Innovation Group, Samsung Electronics\\
}

\vspace{2mm}

\noindent{\emph{Contact: \url{xx.xing@samsung.com}}}

\end{center}



\paragraph{Method Description}

\begin{figure*}[t]
    \centering
    \includegraphics[width=0.8\linewidth]{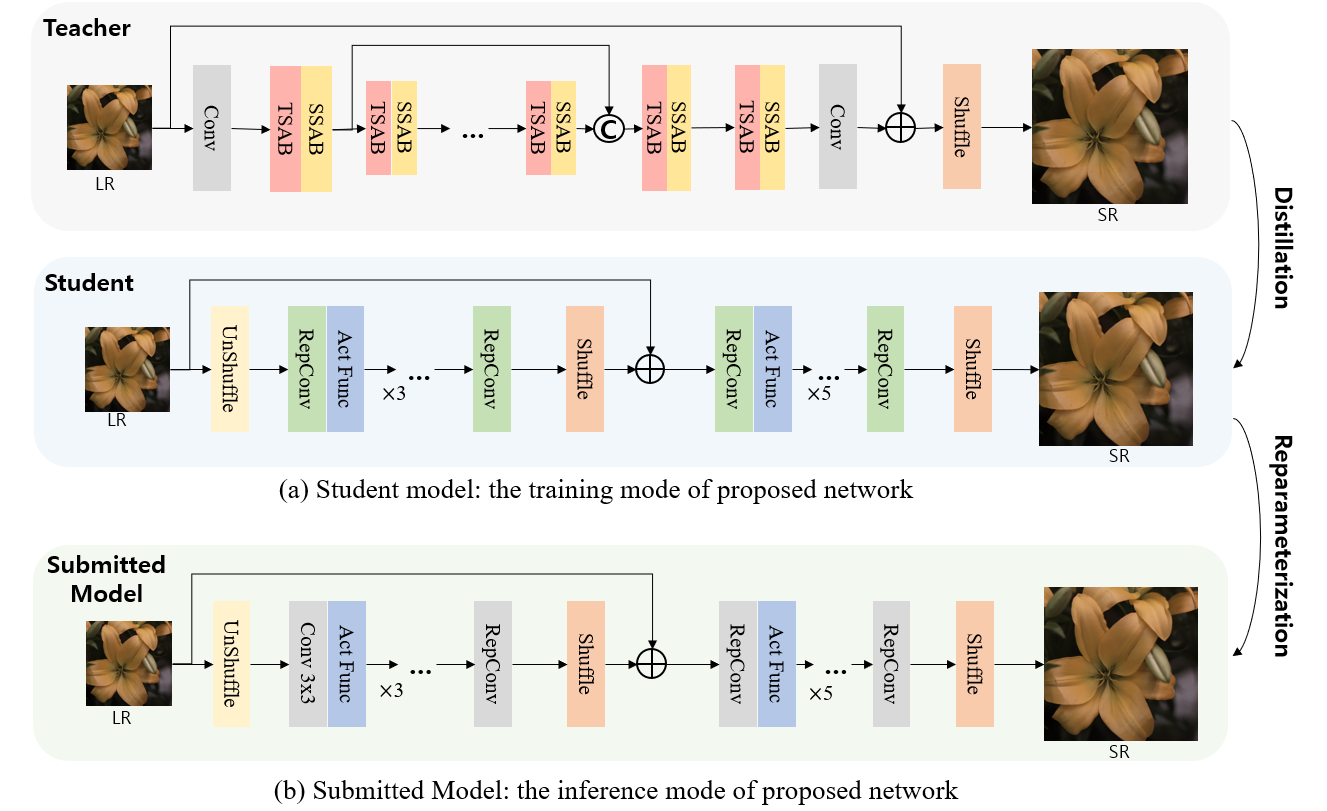}
    \caption{The overall structure of the RawRTSR network, including 0.188M parameters and running at 3.64 ms on the A100 GPU.}
    \label{fig:model1}
\end{figure*}

In the RAW Super Resolution Challenge, we designed our model structure based mainly on CASR~\cite{yoon2024casr}. To meet the parameter and inference time requirements, we adopted the strategy of distillation and parameterization, as shown in Fig.~\ref{fig:model1}. We employ X-Restormerr~\cite{chen2024comparative} as the teacher model, with two architecturally distinct student models trained through knowledge distillation from this shared teacher. Specifically, during the training phase, the student model incorporates re-parameterized convolution blocks. For deployment, these multi-branch convolutional components are structurally converted into a unified convolution layer through parameter fusion, yielding the final submitted model. The specifications of the submitted model are detailed in Table~\ref{tab:samsung_sr_details}

\begin{table*}[t]
    \centering
    \resizebox{\textwidth}{!}{
    \begin{tabular}{c|c|c|c|c|c|c|c|c}
       Type & Model& Input & Training Time & Train E2E & Extra Data & \# Params. (M) & Inference Time & GPU   \\
        \hline
       Efficient Model& RawRTSR~\ref{fig:model1} & $512\times512\times4$ & 24h & Yes & No & 0.19   & 4.45 ms & A100 \\
       General Model  &RawRTSR-L~\ref{fig:model2} & $512\times512\times4$ & 24h & Yes & No & 0.26  & 4.44 ms  & A100\\
    \end{tabular}
    }
    \caption{Training and Testing details of Samsung AI solution.}
    \label{tab:samsung_sr_details}
\end{table*}

\paragraph{Model Framework}
The Raw Super Resolution task encompasses denoising, super-resolution, and blur quality degradation. We decompose these three objectives into two fundamental processes: denoising and detail enhancement. Both two networks employ a straight-through architecture integrating both denoising and detail enhancement modules. 

\begin{figure}[t]
    \centering
    \includegraphics[width=0.98\linewidth]{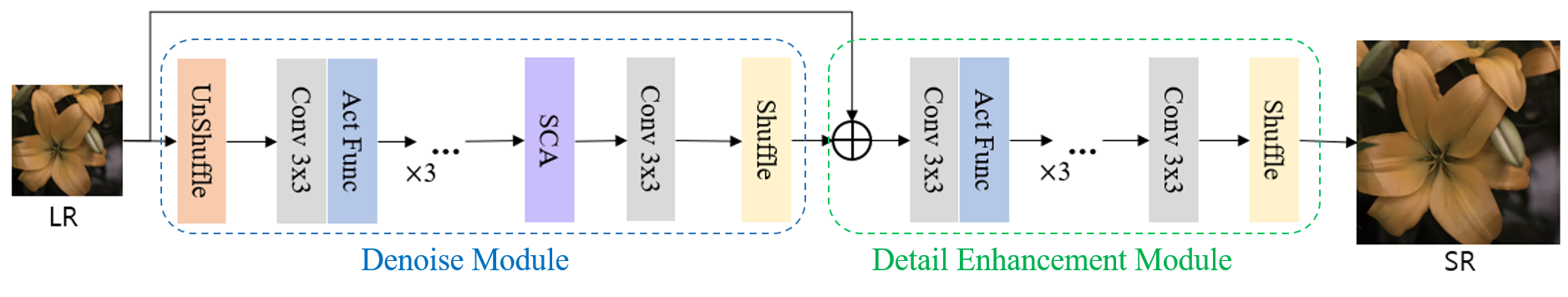}
    \caption{The overall structure of the RawRTSR-L network: 0.311M parameters and running at 4.44 ms on the A100 GPU.}
    \label{fig:model2}
\end{figure}

\begin{itemize}
\item Efficient model, RawRTSR: As shown in Fig.~\ref{fig:model1}, includes denoising module and detail enhancement module. 
Denoising Module first reduces image resolution via unPixelshuffle downsampling to effectively capture global information critical for noise removal.  It processes features through four convolutions for denoising, followed by resolution restoration to match the original input size through upsampling. Detail Enhancement Module employs five convolutions to recover fine textures. To prevent excessive detail loss during denoising, we explicitly incorporate residual connections from the original input. The final output is upscaled through pixel shuffle operations to achieve super-resolution reconstruction. The network maintains a maximum feature channel number of 48 throughout both modules to balance performance and complexity.
\item General Model, RawRTSR-L: As shown in Fig.~\ref{fig:model2}, different from the RawRTSR, we increased the number of feature channels from 48 to 64 to enhance representational capacity. However, to prevent potential information redundancy during the denoising stage resulting from this channel expansion, we incorporated a channel attention mechanism to adaptively recalibrate feature responses.
\end{itemize}

\paragraph{Implementation Details}

We use two synthetic degradation methods to obtain low quality (LQ) images:

\begin{itemize}
    \item Randomly add noise and blur multiple times at RAW domain.
    \item Convert the RAW image to RGB to add motion blur and noise and then back to RAW.
\end{itemize}

We trained our model in three steps. All steps use the PyTorch framework and A100 GPU.

\begin{itemize}
\item First, we train the teacher and student models separately. The LQ patches were cropped from
synthetically degraded LQ images with $256 \times 256$ sizes. AdamW~\cite{decoupled} optimizer($\beta_1=0.9$, $\beta_2=0.999$, weight decay $0.0001$) 
was used with a learning rate of $0.0005$. The total number of epochs was set at $800$. We use the l1 loss.
\item In the second step, the model was initialized with the weights trained in the first step, and feature distillation was used. The total number of epochs was set at $800$. In this step, the initial learning rate was set to $0.00005$ and we use L2 loss.
\item  In the third stage, the model was initialized
using the weights trained in the previous step. The LR patches were cropped from synthetically degraded LQ images with $512 \times 512$ sizes.  
\end{itemize}

In order to ensure that the model parameters and inference time meet the requirements, the final submitted model is the student model after reparameterization.

\begin{figure*}[t]
    \centering
    \includegraphics[width=\textwidth]{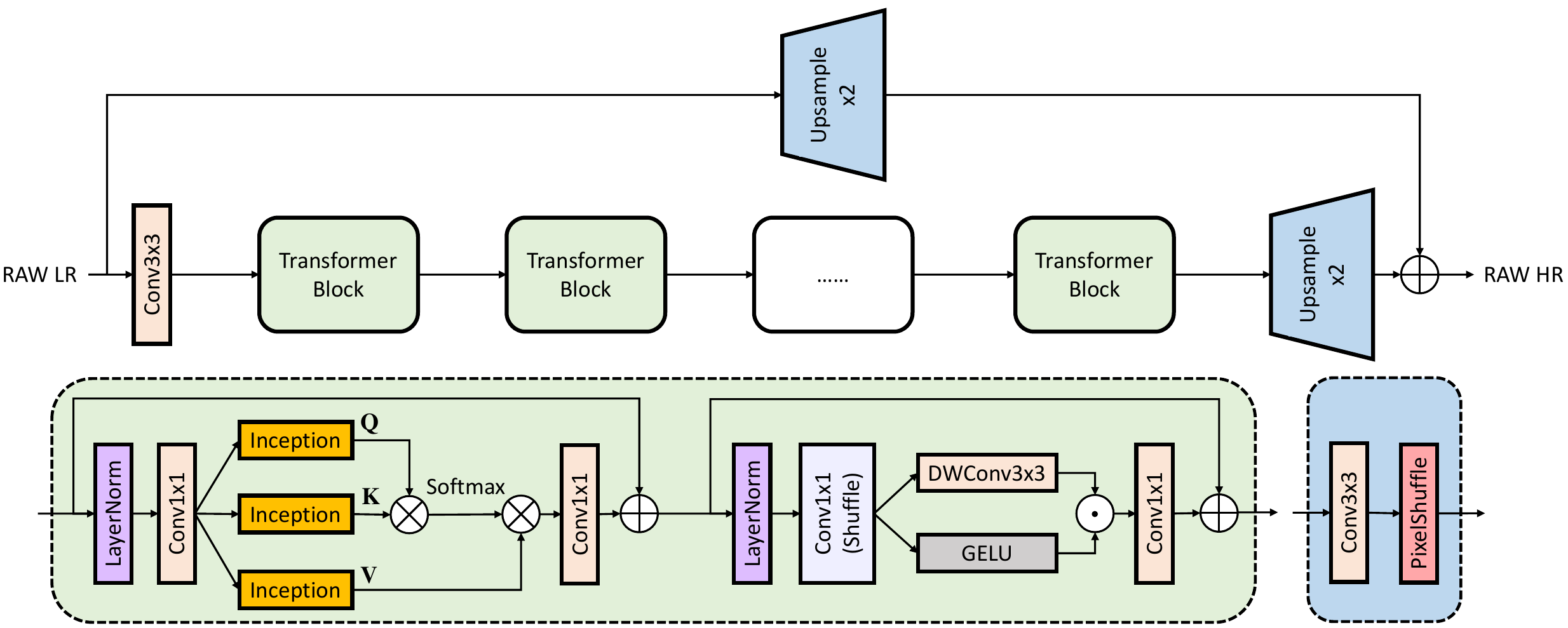}
    \caption{Team USTC framework for RAW image super resolution.}
    \label{visual}
\end{figure*}

\subsection{Streamlined Transformer Network for Real-Time Raw Image Super Resolution}
\label{sec:ustc}

\begin{center}

\vspace{2mm}
\noindent\emph{\textbf{Team USTC-VIDAR}}
\vspace{2mm}

\noindent\emph{Tianyu Zhang\textsuperscript{1}\thanks{Equal contribution.}, 
    Xin Luo\textsuperscript{1}\footnotemark[1], 
    Yeda Chen\textsuperscript{2}, 
    Dong Liu\textsuperscript{1} \\}

\vspace{2mm}

\noindent\emph{
    \textsuperscript{1} University of Science and Technology of China, Hefei, China \\
    \textsuperscript{2} Shanghai Shuangshen Information Technology Co., Ltd.}

\vspace{2mm}

\noindent{\emph{Contact: \url{zhangtianyu@mail.ustc.edu.cn}}}

\end{center}


\paragraph{Method Description}
The overall framework of our method is illustrated in \cref{visual}. It is a streamlined version of RBSFormer \cite{jiang2024rbsformer}, designed for efficient processing. The main branch consists of a 3$\times$3 convolution, $N$ cascaded transformer blocks, and an upsample block, while the residual branch contains only an upsample block. Each upsample block employs a 3$\times$3 convolution followed by a PixelShuffle operation \cite{shi2016real} to upscale features by 2.

The computational complexity of transformers primarily arises from self-attention modules and feed-forward networks, where feature maps are expanded for queries (Q), keys (K), and values (V) or projected for MLP selection. To mitigate these inefficiencies, we incorporate InceptionNeXt \cite{yu2024inceptionnext}, which leverages partial convolution and depth-wise convolution for efficient spatial feature extraction during Q, K, and V projection. For the feed-forward networks, we adopt ShuffleNet \cite{zhang2018shufflenet} with $G$ channel groups to reduce input projection parameters while maintaining cross-channel communication. Inspired by \cite{jiang2024rbsformer, ma2024rewrite}, we further streamline the output projection using element-wise multiplication with a depth-wise convolution gate. We set $N=8$ and $G=4$.

Our method is trained on the NTIRE 2025 RAW Image Super Resolution Challenge dataset, with the degradation pipeline following \cite{conde2024bsraw}. For a 4-channel, 1MP RGGB RAW image, our model requires 519.72 GFLOPS and contains 1.94M parameters, aligning with the challenge’s constraints (0.2M–2M parameters). On an NVIDIA RTX 3090, a forward pass for a full-resolution image takes 96 ms.

\paragraph{Implementation Details}
The model is trained exclusively on the dataset provided by the challenge organizers, which consists of over 1,000 RAW images captured by various DSLR camera sensors. To enhance diversity, the dataset is augmented using random horizontal flips, vertical flips, and transpositions. Degraded images are simulated using the BSRAW degradation pipeline with additional PSF kernels~\cite{hradivs2015convolutional}.

The training process consists of two stages. In the first stage, the model is trained for 300k steps using a batch size of 8 and a patch size of 192. The learning rate is decayed from \(2 \times 10^{-4}\) to \(10^{-6}\), and training takes approximately 12 hours on NVIDIA RTX 3090 GPUs. In the second stage, the batch size is increased to 64, and the patch size is set to 256. The model is trained for 147k steps with the learning rate decaying from \(10^{-4}\) to \(10^{-6}\), requiring around 31 hours on A800 GPUs.  

For both stages, we use the Adam optimizer with default hyperparameters. The loss function is a combination of Charbonnier loss and a Frequency loss~\cite{mao2023intriguing}, with the latter assigned a weight of 0.5.

\begin{figure*}[t]
    \centering
    \includegraphics[width=0.77\textwidth]{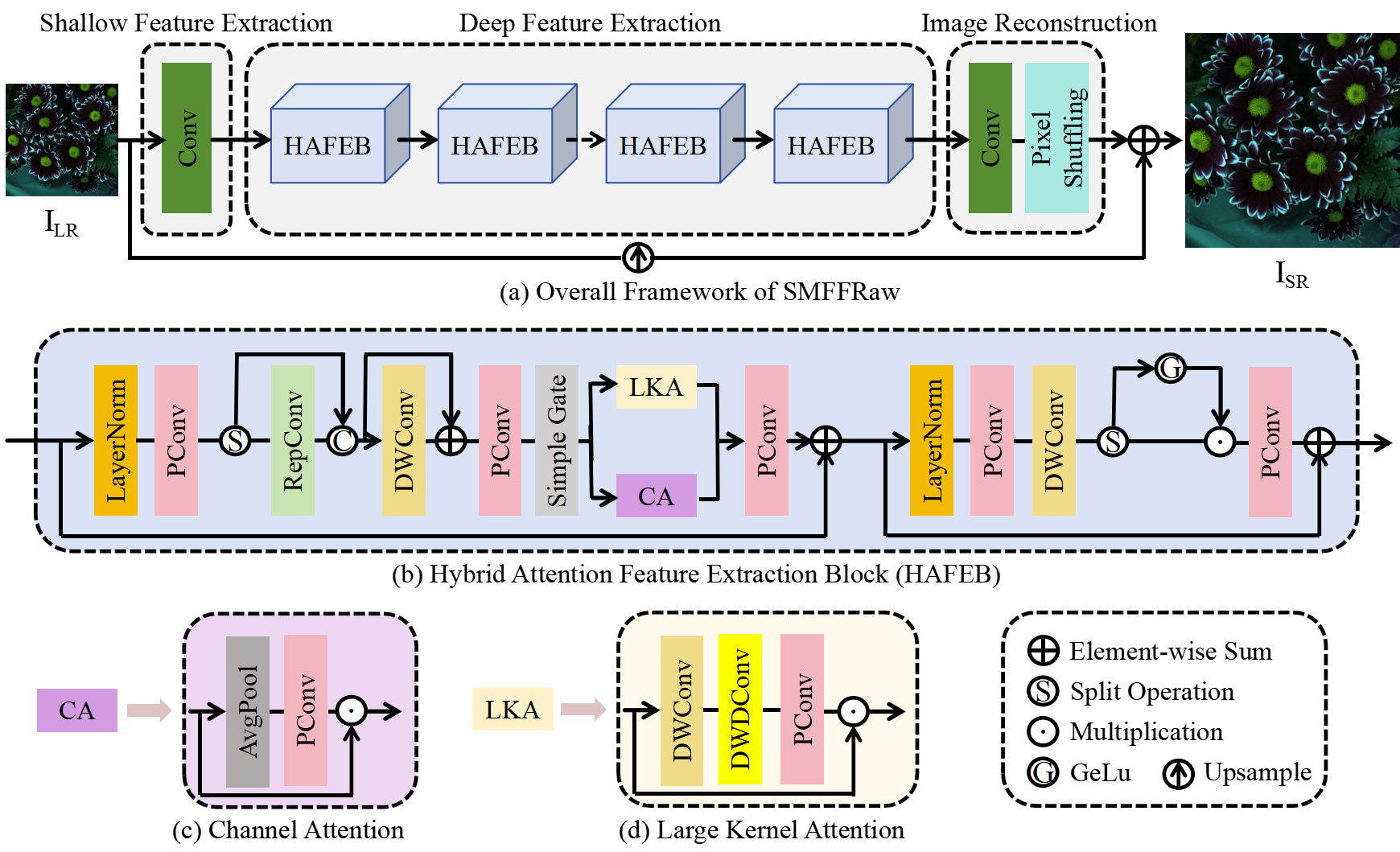}
        \caption{The proposed network architecture of SMFFRaw by Team XJTU. (a) Overall framework. (b) Hybrid Attention Feature Extraction Block (HAFEB). (c) Channel Attention (CA). (d) Large Kernel Attention (LKA).}
    \label{fig:my_diagram}
\end{figure*}

\subsection{SMFFRaw: Simplified Multi-Level Feature Fusion Network for RAW Image Super-Resolution}
\label{sec:xjtusr}

\begin{center}

\vspace{2mm}
\noindent\emph{\textbf{Team XJTU}}
\vspace{2mm}

\author{Li Pang\textsuperscript{1,2*},
Yuhang Yang\textsuperscript{1,2*},
Hongzhong Wang\textsuperscript{1,2*},
Xiangyong Cao\textsuperscript{1,2$\dag$} \\
\textsuperscript{1} School of Computer Science and Technology, Xi’an Jiaotong University, Xi’an, 710049, China.\\
\textsuperscript{2} Ministry of Education Key Laboratory of Intelligent Networks and Network Security
}

\vspace{2mm}

\noindent{\emph{Contact: \url{caoxiangyong@mail.xjtu.edu.cn}}}

\end{center}



While existing RAW image super-resolution methods have achieved remarkable restoration quality, their reliance on over-parameterized architectures leads to computational overhead, rendering them impractical for real-time processing. To bridge this gap, we propose a simplified multi-level feature fusion network for RAW image super-resolution (SMFFRaw), a computationally efficient network based on MFFSSR~\cite{li2024multi}. By employing a novel iterative training strategy which gradually enhances the overall performance of the model, our method achieves 42.208/42.628 dB PSNR on the validation set with only 0.182M/1.994M parameters.

\paragraph{Method Description}

The specific network architecture of SMFFRaw is shown in Figure~\ref{fig:my_diagram}, which consists of three main components including shallow feature extraction, deep feature extraction and reconstruction. More details are provided in the following.

\textbf{Shallow Feature Extraction:} 
Given a degraded input image \(I_{LR}\), a simple \(3 \times 3\) convolutional operation is used to extract shallow features \(F_{0}\).

\textbf{Deep Feature Extraction:} 
Deep features are extracted using a sequential of Hybrid Attention Feature Extraction Block 
(HAFEB) modules, each of which consists of operations such as Point-wise Convolution (Pconv), Depth-wise Convolution (DWconv), Reparameterized Convolution (RepConv), Channel Attention (CA) and Large Kernel Attention (LKA) as demonstrated in Figure~\ref{fig:my_diagram}(b). No reparameterization is applied during the inference phase.

\textbf{Reconstruction:} 
In the reconstruction stage, the feature map is first upsampled using a \(3 \times 3\) convolutional layer with pixel shuffle~\cite{shi2016real}, and then added to the bilinearly interpolated input to produce the final result \(I_{SR}\). This design reduces training complexity while enhancing the network's SR performance.

\begin{table*}[!ht]
    \centering
    \renewcommand{\arraystretch}{0.9}
    \setlength{\tabcolsep}{5pt}
    \small
    \begin{tabular}{lcccccc}
        \toprule
        Input & Training Time & Train E2E & Extra Data & \# Params (M) & GPU \\
        \midrule
        $(512,512,4)$ & ~99h & Yes & No & 0.182 & RTX 4090 \\
        $(512,512,4)$ & ~174h & Yes & No & 1.994 & RTX 4090 \\
        \bottomrule
    \end{tabular}
    \caption{Training details of SMFFRaw showing input dimensions, computational requirements, and model complexity.}
    \label{tab:xjtu_training_details}
\end{table*}

\begin{table*}[!ht]
    \centering
    \renewcommand{\arraystretch}{0.9}
    \setlength{\tabcolsep}{5pt}    
    \small
    \begin{tabular}{ccccccccc}
        \toprule
        Training Phase & Mixup & Downsample & Noise & Blur & Patch Size & Batch Size & Iterations & Training Loss \\
        \midrule
        Phase 1 & $\checkmark$ & $\checkmark$ & - & - & 512 & 8 & 372K & Charbonnier L1+Frequency \\
        Phase 2 & $\checkmark$ & $\checkmark$ & $\checkmark$ & - & 512 & 8 & 372K & Charbonnier L1+Frequency \\
        Phase 3 & $\checkmark$ & $\checkmark$ & $\checkmark$ & $\checkmark$ & 512 & 8 & 372K & Charbonnier L1+Frequency \\
        Phase 4 & - & $\checkmark$ & $\checkmark$ & $\checkmark$ & 1024 & 4 & 266K & Charbonnier L1+Frequency \\
        Phase 5 & - & $\checkmark$ & $\checkmark$ & $\checkmark$ & 1024 & 4 & 266K & MSE+Frequency \\
        \bottomrule
    \end{tabular}
    \caption{Illustration of the progressive training strategy of SMFFRaw, including augmentation techniques (i.e., mixup), degradation process (i.e., downsample, noise, blur), training patch size, batch size, training iterations and loss functions across different phases.}
    \label{tab:training_phases}
\end{table*}

\begin{table*}[!ht]
    \centering
    \renewcommand{\arraystretch}{0.9}
    \setlength{\tabcolsep}{5pt}
    \small
    \begin{tabular}{l c c c c c}
        \toprule
        Method & \multicolumn{1}{c}{Validation 1MP} & \multicolumn{1}{c}{Test 1MP} & \# Params. (M) & FLOPs (G) & Runtime (ms) \\
        \midrule
        SMFFRaw-Small & 42.208/0.983 & Unknown & 0.182 & 43.58 & 3.5 \\
        SMFFRaw-Large & 42.628/0.984 & Unknown & 1.994 & 478.75 & 3.8 \\
        \bottomrule
    \end{tabular}
    
    \caption{Performance of Team XJTU SMFFRaw solutions, including PSNR/SSIM on the validation set (40 images), parameter number, FLOPs and runtime.}
    \label{tab:xjtu_sr_result}
\end{table*}

\paragraph{Implementation Details}

The model is trained solely using the dataset provided by the challenge organizers. Our training pipeline consists of five stages (see Table~\ref{tab:training_phases}). To increase data diversity, we apply common augmentations (rotation, flipping) along with mixup~\cite{zhang2022swinfir}. Degradation pipeline proposed by BSRAW~\cite{conde2024bsraw} is used to generate RAW-degraded image pairs. The model is trained using the Adam optimizer~\cite{kingma2014adam} with an initial learning rate of 1e-3, which decays to 1e-6 using Cosine Annealing. The network is trained using a combination of Charbonnier loss~\cite{Wang_2019_CVPR_Workshops} and frequency loss~\cite{Jiang_2024_CVPR} for the first four stages, similar to RBSFormer~\cite{jiang2024rbsformer}, and a combination of MSE and frequency loss for the final stage. All experiments are conducted employing PyTorch on RTX 4090 GPUs. The model efficiency results are presented in Table~\ref{tab:xjtu_sr_result} and more training details are provided in Table~\ref{tab:xjtu_training_details}.
\subsection{An Enhanced Transformer Network for Raw Image Super-Resolution}
\label{sec:egroup}

\begin{center}

\vspace{2mm}
\noindent\emph{\textbf{Team EGROUP}}
\vspace{2mm}

\author{Ruixuan Jiang,
Senyan Xu,
Siyuan Jiang,
Xueyang Fu,
Zheng-Jun Zha
\\
University of Science and Technology of China
}

\vspace{2mm}

\noindent{\emph{Contact: \url{rxjiang21@gmail.com}}}

\end{center}


\begin{figure*}[t]
    \centering
    \includegraphics[width=\textwidth]{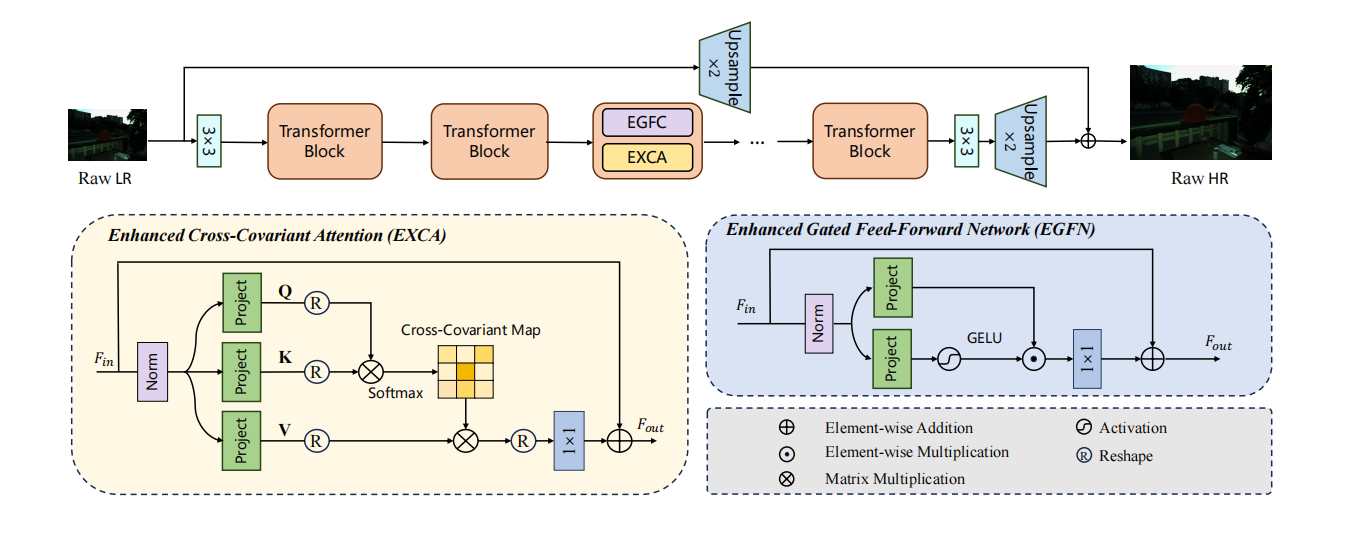}
    \caption{The architecture of the  RBSFormer~\cite{jiang2024} used by Team EGROUP for RAW image super-resolution.}
    \label{fig:egroup}
\end{figure*}

\paragraph{Method Description}
Our approach leverages the RBSFormer\cite{jiang2024} architecture to directly process RAW images for super-resolution tasks. By operating in the RAW domain rather than on sRGB images processed by ISP, we avoid the complexities of non-linear transformations that make degradation modeling challenging.

We maintain the three-component structure of RBSFormer\cite{jiang2024}. The overall pipeline can be described as follows:

Firstly, given a raw low-resolution image with degradation $I_{LR} \in \mathbb{R}^{H\times W\times 4}$, we apply shallow feature extraction:
\begin{equation}
F_s = \text{Conv}_{3\times 3}(I_{LR})
\end{equation}

Next, we use transformer blocks to extract deep features:
\begin{equation}
F_i = \mathcal{H}_{tb_i}(F_{i-1}), i = 1, 2, \ldots, K
\end{equation}
\begin{equation}
F_d = \text{Conv}_{3\times 3}(F_K)
\end{equation}

Finally, we reconstruct the HR image by aggregating the features:
\begin{equation}
I_{HR} = \mathcal{H}_{rec}(I_{LR}, F_d) = \text{Up}(F_s + F_d)
\end{equation}

We utilized the official training dataset provided by the organizers. Our focus has been on optimizing the training process through careful data augmentation in the RAW domain, implementing random noise and blur degradation patterns.

Our evaluation shows competitive results on the validation set, achieving 42.54 dB PSNR and 0.98 SSIM with only 3.3M parameters and reasonable computational requirements, making our approach both effective and efficient for RAW image super-resolution.

\paragraph{Implementation Details}

AdamW optimizer with $\beta_1$ and $\beta_2$ equal to 0.9 and 0.999 is used to train our model. The initial learning rate is set as $7 \times 10^{-4}$. We adopt the cosine annealing strategy to train the models, where the learning rate gradually decreases from the initial learning rate to $1 \times 10^{-6}$. All experiments are implemented by PyTorch 1.11.0 with two NVIDIA 4090 GPUs. We used batch-size of 8 and crop-size of 192.We utilized the official training dataset provided by the organizers.We trained with L1 loss for 100k iterations, then fine-tuned with FFT loss for 20k iterations. %
\subsection{A fast neural network to do super-resolution based on NAFSSR}
\label{sec:nju}

\begin{center}

\vspace{2mm}
\noindent\emph{\textbf{Team NJU}}
\vspace{2mm}

\noindent\emph{Xiang Yu, Guanlan Hong}

\vspace{2mm}

\noindent\emph{Nanjing University}

\vspace{2mm}

\noindent{\emph{Contact: \url{a2417831512@gmail.com}}}

\end{center}


Team NJU RSR proposed a CNN framework for raw image super-resolution, with a design based on NAFBlock propsed in NAFSSR ~\cite{chu2022nafssrstereoimagesuperresolution}, adopting reparameterization during inference, fusing the additional parameters of Batch Normalization into the CNN layer before to achieve efficient inference.

\paragraph{Method Description}

The solution used the data provided in the NTIRE 2025 RAW Image Super Resolution challenges, with the degradation pipeline described in \cite{conde2024bsraw}.

\begin{figure*}[t]
    \centering
    \includegraphics[width=0.6\linewidth]{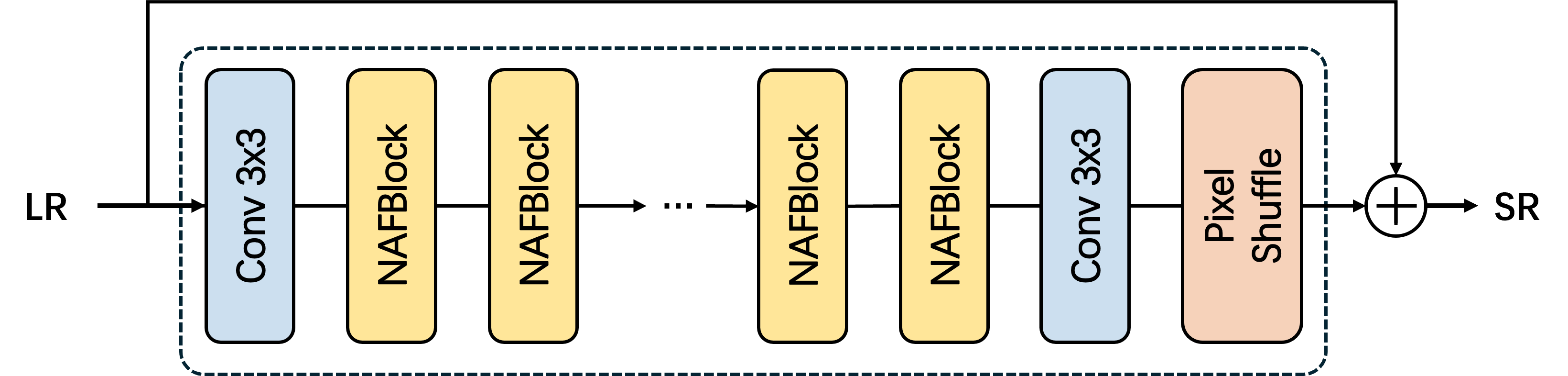}
    \caption{NAFBN proposed by Team NJU RSR.}
    \label{fig:NAFBN}
\end{figure*}

\begin{figure*}[t]
    \centering
    \includegraphics[width=0.6\linewidth]{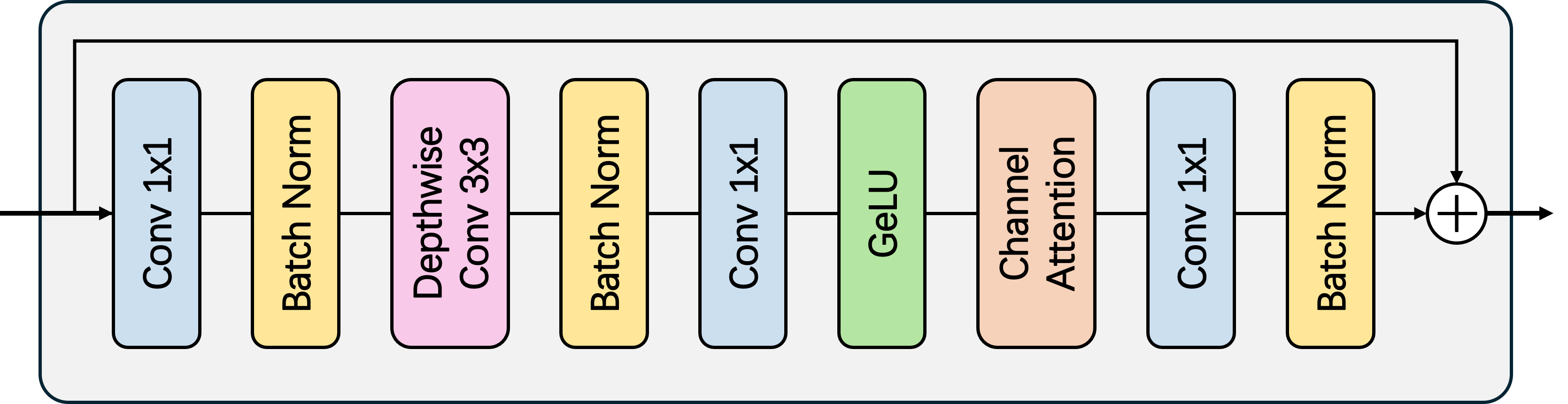}
    \caption{Adopted NAFBlock used by Team NJU RSR.}
    \label{fig:NAFBlock}
\end{figure*}

The architectural configuration of the Team NJU RSR is depicted in Fig \ref{fig:NAFBN}. In the proposed method, a NAFSSR-based design is redesigned to handle single view picture. The team adopted the NAFBlock by redesign the SimpleGate component with a CNN layer and GeLU activation function and removing FFN component to constrain the parameters of the block to build a light-weight model. The adopted NAFBlock is shown in Fig \ref{fig:NAFBlock}. To make the training more easily and inference more efficiently, they replace Layer Normalization with Batch Normalization, which can be fused into the near CNN layer. 

For a 4-channel RGGB RAW image patch of size $256\times256$, the computational cost of the NAFBN amounts to $11.90$ GFLOPS, being characterized by a number of 189K trainable parameters after fusing Batch Normalization layers. On a consumer-grade gaming GPU, the NVIDIA RTX3090, the forward pass needed of a 4-channel RGGB RAW image of size $256\times256$ estimation needs $7.19$ ms after simply fusing Batch Normalization layers and $5.19$ ms using half precision with almost no performance drop comparing with $9.49$ ms without fusion operation.

\paragraph{Implementation Details}

The experiments are based on the Pytorch framework for implementation, with the experiments being performed on single vGPU-32 device provided by the platform \href{https://www.autodl.com/home}{AutoDL}. The NAFBN model uses 12 blocks of NAFBlocks of width 48. The training procedure is based on the Adamw optimizer with the decay parameters $\beta_1 = 0.9$ and $\beta_2 = 0.99$. The momentum of Batch Normalization is set to 0.03 to avoid the performance drop due to limited batch size. The initial learning rate is $1\times 10^{-3}$ and changes with Cosine Annealing scheme to $1\times 10^{-6}$, with the training procedure covering 50K iterations in a time-frame of around 7 hours.

The data augmentation includes random patch cropping of size $32 \times 32$, random white balance, applied in the unit interval normalized images, random horizontal or vetical flips, random right-angle rotations and exposure adjustment by linearly scaling the images with the factor tuned to the [-0.1, 0.1] interval, which are all set with probability of $0.5$ to be applied. To achieve a better performance on relatively lower resolution images, random dowsample with AvePool2d and bicubic interpolation is added during each patch cropping procedure with the probablity of $0.3$. The training objective is based on the $L_1$ loss.

\subsection{A efficient neural network baseline report using Mamba}
\label{sec:tysl}

\begin{center}

\vspace{2mm}
\noindent\emph{\textbf{Team TYSL}}
\vspace{2mm}

\noindent\emph{Minmin Yi, Yuanjia Chen, Liwen Zhang, Zijie Jin}

\vspace{2mm}

\noindent\emph{E-surfing Vision Technology Co., Ltd, China}

\vspace{2mm}

\noindent{\emph{Contact: \url{441095434@qq.com}}}

\end{center}

We implemented the MambaIRv2 method on the raw data with the aim of providing a baseline from a different perspective for the competition.

\paragraph{Method Description}
\begin{figure}[h]
	\centering
	\includegraphics[width=0.45\textwidth]{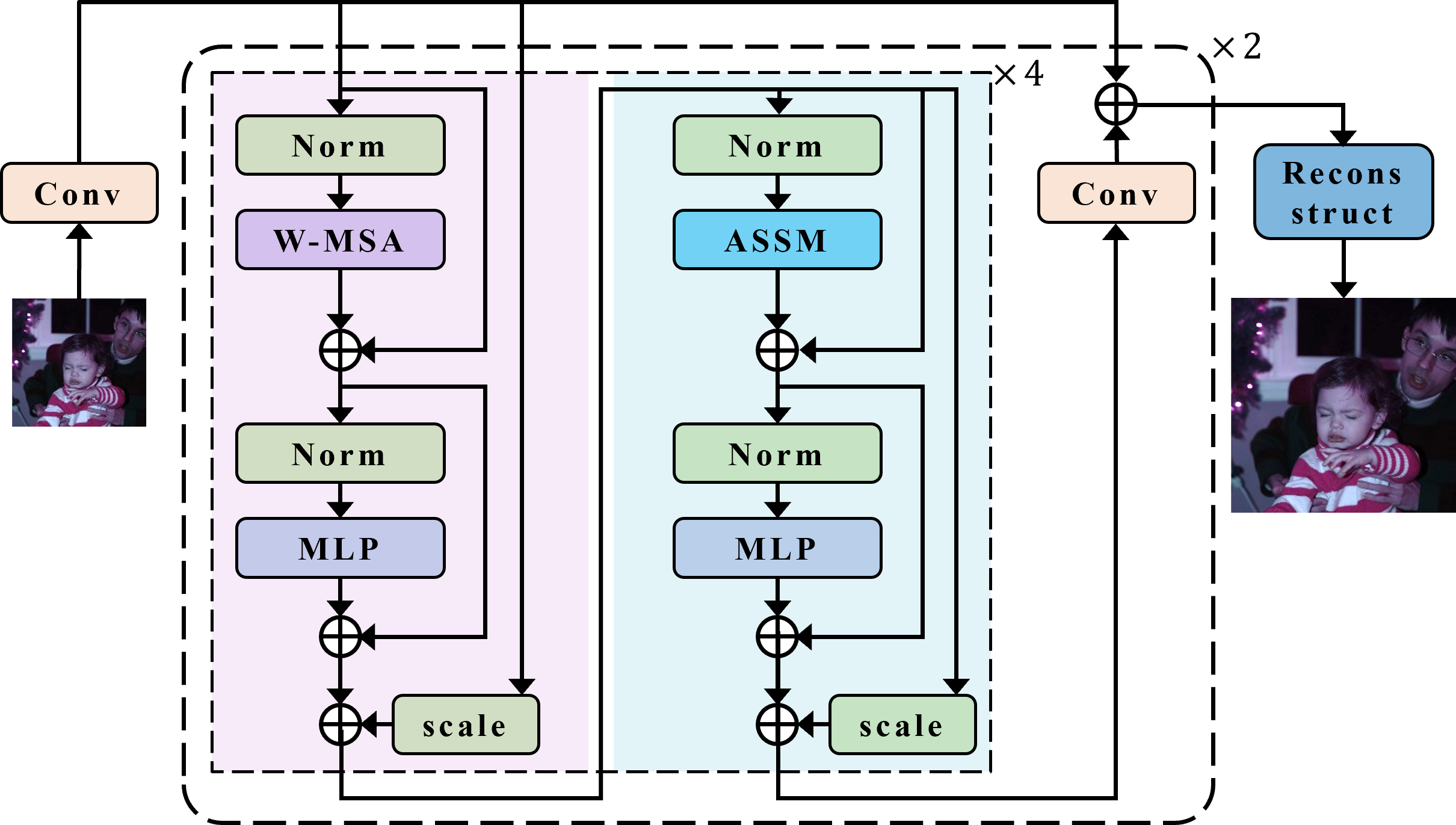}
	\caption{MambaIRv2 structure}
	\label{tysl_fig1}
\end{figure}

We implemented the MambaIRv2{\cite{guo2024mambairv2}} method on raw data with the aim of providing a baseline from a different perspective for the competition. This method did not use any additional data, and we simplified its architecture to obtain a lightweight model with a size of less than 0.2m. The architecture of this model is shown in Figure \ref{tysl_fig1}, where embedding dim = 32, m = 4, and n = 2. We chose the Mamba model because of its great potential for lightweighting and because no one had ever used it on raw data before.
In addition, we have done a lot of work on downsampling. We have tried many downsampling methods, including 1. directly performing bicubic downsampling on each channel, 2. using AvgPool2D as provided in the competition, and 3. using bicubic downsampling with bias, as shown in Figure 2. However, among these methods, AvgPool2D performs much better than the others. Since we do not know whether the test set images in the competition are synthetic or real data, if they are synthetic data, the downsampling in the synthesis method will have a significant impact.

\paragraph{Implementation Details}
\begin{figure}[h]
	\centering
	\includegraphics[width=0.35\textwidth]{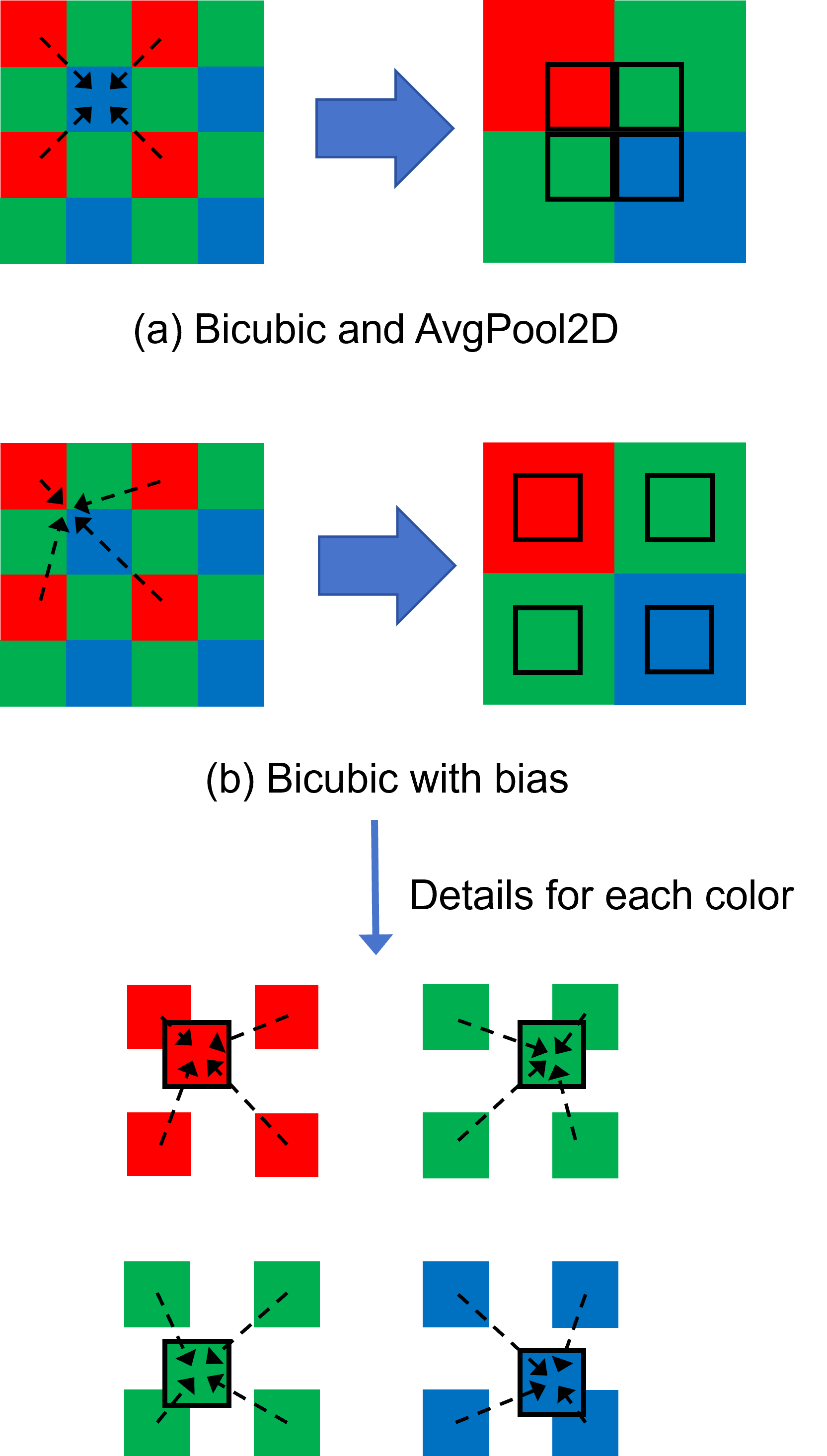}
	\caption{Downsampling structure}
	\label{tysl_fig2}
\end{figure}

We propose a new downsampling method based on the intuition that the central pixel value of a region is closer to the region's average pixel value. Taking the smallest 4×4 unit as an example, after downsampling, it transforms into a 2×2 unit with an RGGB arrangement. In this structure, the red pixel at the top-left corner after downsampling represents the average value of the top-left 2×2 pixels before downsampling, as shown in Figure \ref{tysl_fig2}(b).  
If we directly apply bicubic downsampling or `avgpool2d` to the red channel of the original image, the resulting value should approximate the corresponding position in the original 4×4 image, specifically the pixel at coordinate (1,1), as illustrated in Figure \ref{tysl_fig2}(a). Notably, this value ends up at the bottom-right of the red pixel in the downsampled 2×2 image rather than at its center.  
Therefore, if we perform bicubic interpolation using the nearest 16 points, with the interpolation point positioned at the center of the downsampled red pixel, we believe this approach achieves more precise downsampling.

As for other settings, we conducted our experiments on the server with several A100 GPUs, using the PyTorch framework. The batch size was set to 64, and the learning rate was configured as 8e-4, resulting in a total training time of approximately 26 hours. We identified that the primary factor contributing to the slow training speed was the prolonged image degradation process. Utilizing the provided training set, we adhered to the given degradation pipeline, except for the downsampling step, and did not apply any image enhancement. 

 %
\subsection{RepRawSR: Accelerating Raw Image Super-Resolution with Reparameterization}
\label{sec:eifflow}

\begin{center}

\vspace{2mm}
\noindent\emph{\textbf{Team EiffLowCVer}}
\vspace{2mm}

\author{Jiajie Lu\\
Politecnico di Milano

}
\vspace{2mm}
\noindent{\emph{Contact: \url{lujiajie010@126.com}}}

\end{center}


\begin{table}
    \centering
    \begin{tabular}{c @{\hspace{15pt}} c @{\hspace{15pt}} c @{\hspace{15pt}} c}
         Method & FLOPs & Val (Self) & Val  \\
         \hline
         NAFnet-1.9M(baseline)& 9.68G & 40.21 & 41.70 \\
         RepLarge-97k & 24.42G &39.22 & 40.80 \\
         RepTiny-21k  & 5.65G  &39.00 & -- \\
    \end{tabular}
    \caption{Ablation study by Team Team EiffLowCVer. PSNR results on the self-organized validation set (40 images selected from the training set) and the validation set (40 images). Testing of the Tiny-21k model on Codalab was not conducted due to submission limits in the test phase.}
    \label{tab:results}
\end{table}

\paragraph{Method Description}

We designed two lightweight variant of previous SYEnet\cite{Gou_2023_ICCV} for raw image super-resolution, incorporating structural reparameterization and efficient network design. The Large-97k model uses 32 channels as the global feature dimension and adds an additional feature extraction module before the original SYEnet\cite{Gou_2023_ICCV} blocks, resulting in higher PSNR performance. The Tiny-21k model employs 16-channel feature extraction modules, achieving faster processing speeds and minimal computational complexity with only 21k parameters after reparameterization.

We conducted secondary development based on SYEnet and found that common operations, such as increasing the number of processing blocks or simplely enlarging the channel numbers, tend to cause instability during training, resulting in suboptimal performance. To stabilize the training process and achieve better results, we carried out extensive experiments and ultimately trained the following two models on the provided training dataset.

\begin{itemize}
\item RepTiny-21k: The original SYEnet used only one feature extraction module, which limiting feature extraction capability. Our Tiny-21k model increases this to four and introduces skip connections (red arrows) to mitigate gradient vanishing. We set the channel number to 16 for efficiency.With an input of (512,512,4), our model achieves 5.65G FLOPs, 21k parameters, and 39.00 dB PSNR on a self-organized validation set
\item RepLarge-97k: Another approach to enhance the model is increasing the channel width instead of network depth. We set the channel number to 32, used only one Feature Extraction Module, and incorporated the FEBlock—a preprocessing module from SYEnet designed for super-resolution task. This configuration improved PSNR by 0.22 dB on the self-organized validation set. However, due to the wider channels and added FEBlock, the parameters and FLOPs increased to 5 times the original. \\ 

Based on the experimental results, we observe that increasing the number of feature extraction modules while employing skip connections for training stability enables Tiny-21k to achieve an optimal balance between performance and speed.

\end{itemize}

\begin{figure*}[t]
    \centering
    \includegraphics[width=1.0\textwidth]{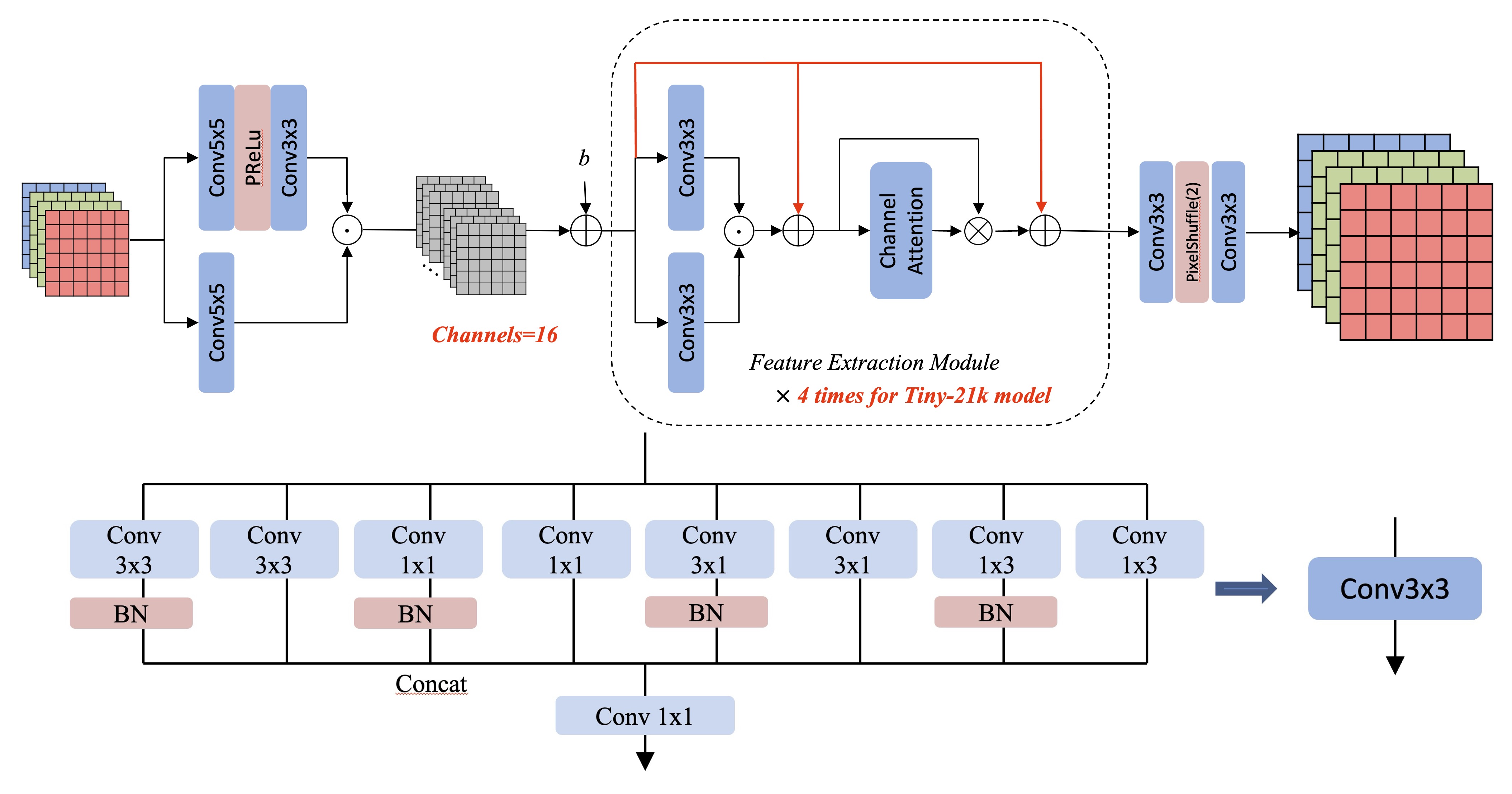}
    \caption{Main branch of RepRawSR proposed by Team EffiLowCVer.}
    \label{fig:EffiLowCVer}
\end{figure*}

\begin{table*}[]
    \centering
    \begin{tabular}{c|c|c|c|c|c}
        Input & Training Time & Train E2E & Extra Data & \# Params. (M) & GPU  \\
        \hline
         (256,256,3) & 22h & Yes & No & 0.021 Million for Tiny-21k & 3090 \\
         (256,256,3) & 26h & Yes & No & 0.097 Million for Large-97k & 3090
    \end{tabular}
    \caption{The overall training details of RepRawSR.}
    \label{tab:eifflow_result}
\end{table*}

\paragraph{Implementation Details}

\begin{itemize}
    \item \textbf{Optimizer and Learning Rate:} We used the Adam optimizer with an initial learning rate of \(8\times10^{-4}\) and adopted CosineAnnealingRestartLR to schedule the learning rate.
    \item \textbf{GPU:} NVIDIA GeForce RTX 3090 24Gb
    \item \textbf{Datasets:} We used only the 1,064 RAW images provided by the organizers as the dataset, with 40 images randomly selected as the validation set during training.
    \item \textbf{Training Time:} 22 hours
    \item \textbf{Training Strategies:} We adopted a multi-stage training strategy. In the first stage, during 100,000 training steps, we randomly cropped 256×256 GT patches, applied random rotation and flipping, and then generated LQ images using the degradation pipeline provided by the organizers as model input. In the second stage, the patch size was increased to 384×384, and training continued for an additional 50,000 steps.
    \item \textbf{Efficiency Optimization Strategies:} To stabilize training, an additional tail generated a second predicted image from intermediate feature maps, which was included in the loss calculation. This branch is removed during inference.
\end{itemize}
 %
\begin{figure*}[t]
    \centering
    \includegraphics[width=0.8\linewidth]{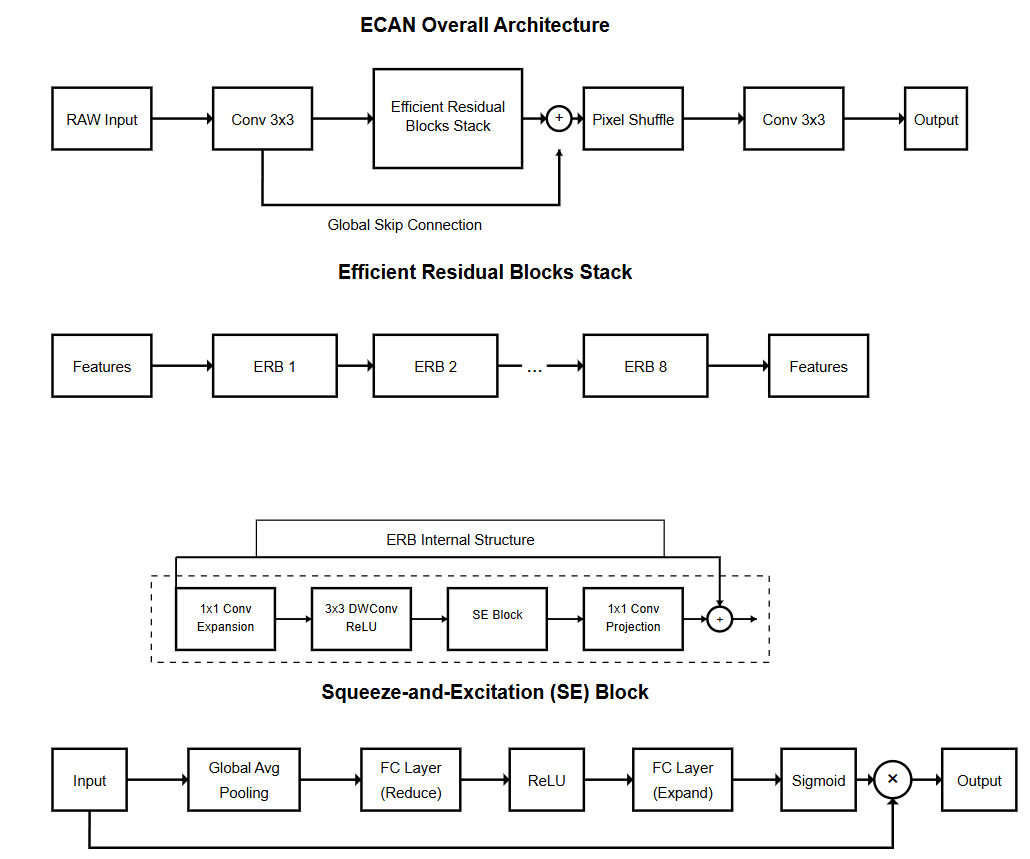} 
    \caption{The overall architecture of the proposed ECAN model, showing the main pipeline and details of the EfficientResidualBlock with the integrated SEBlock.}
    \label{fig:ecan_arch}
\end{figure*}

\subsection{ECAN: Efficient Channel Attention Network for RAW Image Super-Resolution}
\label{sec:cuee}

\begin{center}

\vspace{2mm}
\noindent\emph{\textbf{Team CUEE-MDAP}}
\vspace{2mm}

\noindent\emph{Watchara Ruangsang}

\vspace{2mm}

\noindent\emph{Multimedia Data Analytics and Processing Unit, \\
    Department of Electrical Engineering, Faculty of Engineering, \\
    Chulalongkorn University, Thailand}

\vspace{2mm}

\noindent{\emph{Contact: \url{watchara.knot@gmail.com}}}

\end{center}


\paragraph{Method Description}

Many high-performing super-resolution (SR) models suffer from high computational complexity. As noted in \cite{ruangsang2017efficient}, "many approaches are complex and are difficult to use in practical scenarios." Furthermore, "computational cost and model parameters are the most challenging limitations in real-world applications" \cite{ruangsang2023multi}. ECAN aims to create an "efficient Super-resolution algorithm" \cite{ruangsang2017efficient} for the NTIRE 2025 Efficient Track (<0.2M parameters).

ECAN is a CNN-based model. No external pre-trained models were used. It was trained end-to-end on the NTIRE 2025 RAW training dataset using standard augmentations and a specific degradation pipeline.

The ECAN architecture (Fig.~\ref{fig:ecan_arch}) uses four stages: (1) Shallow Feature Extraction (3x3 Conv on 4-channel RAW input), (2) Deep Feature Extraction (8 `EfficientResidualBlocks` with global skip connection), (3) Upsampling (PixelShuffle), and (4) Reconstruction (3x3 Conv to 4-channel RAW). Each `EfficientResidualBlock` uses an inverted residual structure with depthwise separable convolutions (inspired by MobileNetV2 \cite{sandler2018mobilenetv2}) and a Squeeze-and-Excitation (SE) block \cite{hu2018squeeze} for channel attention. This focus on channel interaction aligns with our group's work on Multi-FusNet \cite{ruangsang2023multi} and channel attention networks \cite{zhang2018rcan}.

ECAN has only \textbf{93,092 parameters} ($\approx$ 0.093M), significantly below the 0.2M limit. The computational cost is estimated at 21.82 GMACs (or 43.65 GFLOPs) for a 512x512x4 input (scaling to 1MP output). Inference time is approximately \textbf{8.25 ms per output megapixel} on an NVIDIA RTX 4090 (measured without AMP). 


\paragraph{Implementation Details}

\begin{itemize}
    \item \textbf{Framework:} PyTorch
    \item \textbf{Optimizer:} AdamW ($\beta_1=0.9, \beta_2=0.999$), weight decay $1 \times 10^{-4}$.
    \item \textbf{Learning Rate:} Initial $4 \times 10^{-4}$, cosine annealing to $1 \times 10^{-7}$.
    \item \textbf{GPU:} NVIDIA RTX 4090 (24GB).
    \item \textbf{Datasets:} NTIRE 2025 RAW training set. Augmentation: Random 90/180/270 rotations, horizontal flips. Degradation: Gaussian Blur ($\sigma \le 4.0, p=0.7$) + Gaussian Noise ($level \le 0.04, p=0.95$). No extra data used.
    \item \textbf{Training Time:} 600 epochs (~1.6 hours).
    \item \textbf{Training Strategies:} End-to-end from scratch, Automatic Mixed Precision (AMP). Input patch size 128x128, batch size 64. L1 Loss. Gradient clipping at 1.0.
\end{itemize}

\begin{figure*}[!ht]
    \centering
    \setlength{\tabcolsep}{0.6pt}
    \begin{tabular}{c c c c c}
         \includegraphics[width=0.20\linewidth]{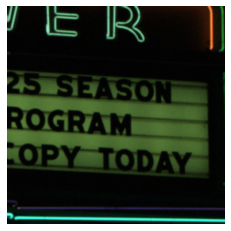} &
         \includegraphics[width=0.20\linewidth]{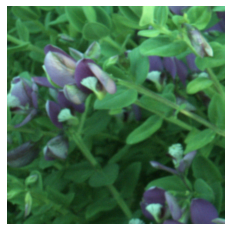} &
         \includegraphics[width=0.20\linewidth]{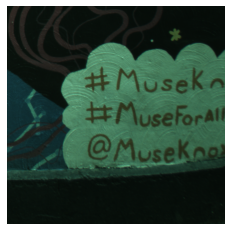} &
         \includegraphics[width=0.20\linewidth]{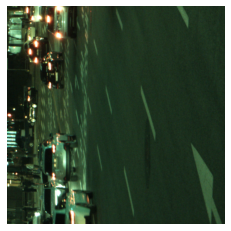} &
         \includegraphics[width=0.20\linewidth]{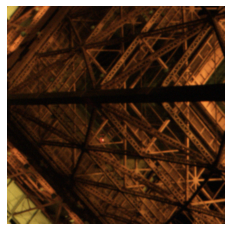} \\
    \end{tabular}
    \caption{RAW image samples from RawIR Dataset.}
    \label{fig:rawir-samples}
\end{figure*}

\newpage

\section{RAW Image Restoration Methods}
\label{sec:irteams}

\subsection{Efficient RAW Image Restoration}
\label{sec:samsungir}

\begin{center}

\vspace{2mm}
\noindent\emph{\textbf{Team SamsungAI}}
\vspace{2mm}

\author{
Xiangyu Kong~$^1$,
Xiaoxia Xing~$^1$, \\
Suejin Han~$^2$,
MinKyu Park~$^2$ \\
$^1$ Samsung R\&D Institute China - Beijing (SRC-B)\\
$^2$ The Department of Camera Innovation Group, Samsung Electronics\\
}
\vspace{2mm}
\noindent{\emph{Contact: \url{xiangyu.kong@samsung.com}}}

\end{center}

\paragraph{Method Description}
In the RAW Restoration Challenge, we designed our model structure based mainly on Nafnet~\cite{chen2022simple}, a lightweight network for RAW image restoration, available in two variants: ERIRNet-S (5M parameters) and ERIRNet-T (0.2M parameters). To satisfy parameter constraints while preserving performance under diverse mobile RAW degradations, we reduced Nafnet parameters and implemented the distillation strategy. X-Restormer~\cite{chen2024comparative} is adopted as the teacher model, with two architecturally distinct student models trained through knowledge distillation from this shared teacher. 

\begin{figure}[ht]
    \centering
    \includegraphics[width=0.98\linewidth]{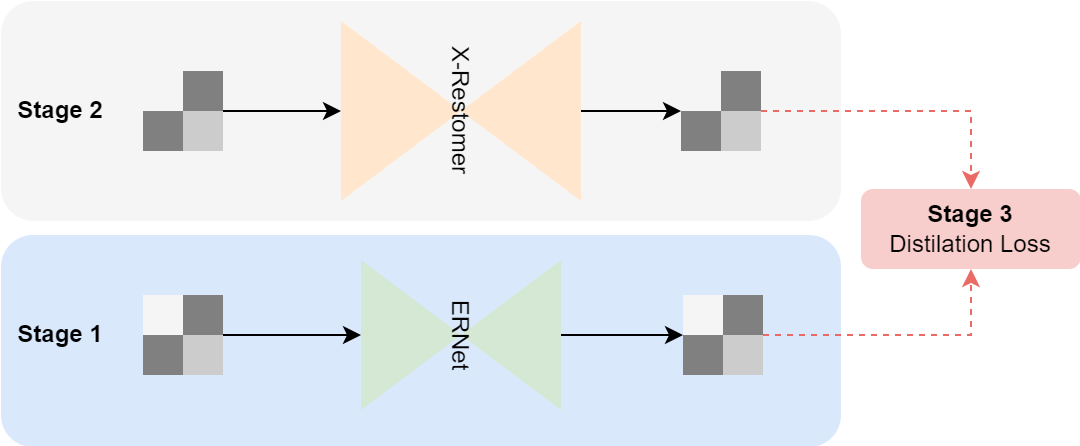}
    \caption{Training Stage Description by Samsung AI}
    \label{fig:samsung_teacher_student}
\end{figure}

\begin{table*}[t]
    \centering

    \begin{tabular}{c|c|c|c|c|c|c|c}
        Model & Input & Training Time & Train E2E & Extra Data &  FLOPs (GMac) & \# Params. (M) & GPU  \\
        \hline
        ERIRNet-S & (512, 512, 4) & 24h & Yes & No &  23.79 & 4.97 & A100\\
        ERIRNet-T & (512, 512, 4) & 26h & Yes & No &  10.98 & 0.19 & A100
    \end{tabular}
    
    \caption{Training details of SamsungAI solution.}
    \label{tab:samsung_train_label}
\end{table*}

\paragraph{Model Framework}
Raw image restoration addresses coupled degradation processes involving noise suppression and blur correction. Methodologically, we select the NafBlock architecture from Nafet~\cite{chen2022simple} as architecturally pivotal components, given their demonstrated SOTA efficacy in joint denoising and deblurring tasks. To achieve parameter efficiency without compromising restoration fidelity, we designed networks based on a narrow-and-deep architectural principle, reducing channel dimensions while preserving layer depth, which resulted in two architecture variants with distinct parameter configurations as shown in Table~\ref{tab:samsung_train_label}:

\begin{itemize}
    \item ERIRNet-S is a simplified version of NAFNet with reduced channel width and fewer encoder-decoder blocks for improved efficiency.
    \item ERIRNet-T further reduces complexity by decreasing the number of blocks and using smaller FFN expansion ratios. It also replaces PixelUnshuffle layers with ConvTranspose, enabling deeper architectures under a strict parameter budget.
\end{itemize}

\begin{figure}
    \centering
    \includegraphics[width=0.98\linewidth]{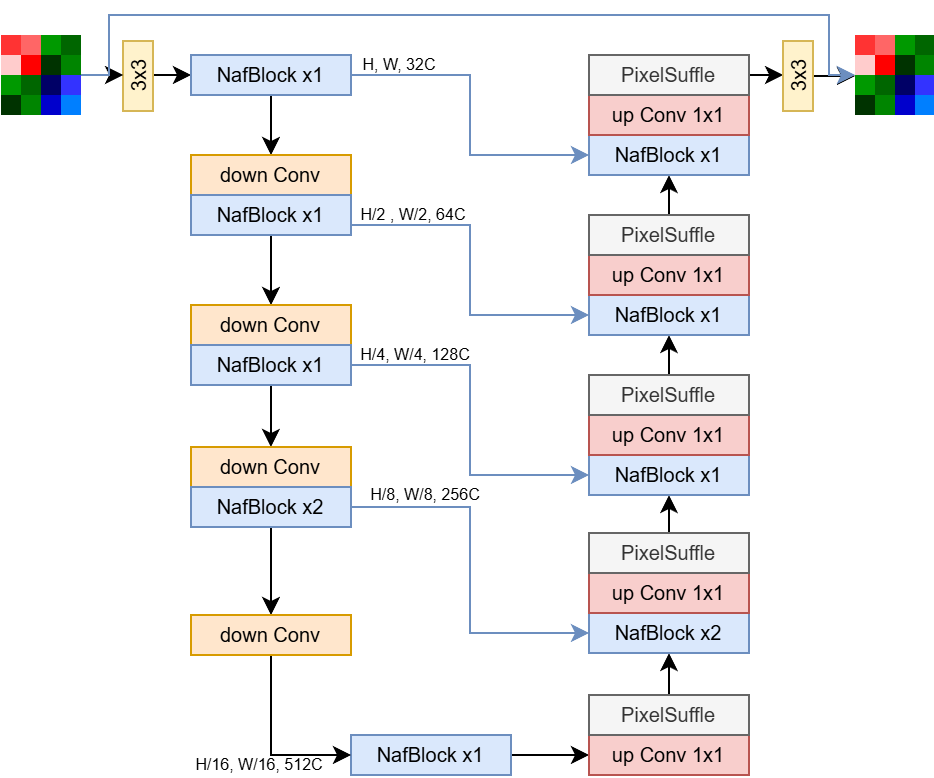}
        \caption{Architectures of ERIRNet-S, with reduced channels and fewer blocks. Proposal by Samsung AI.}
    \label{fig:erin_s}
\end{figure}

\paragraph{Implementation Details}
We trained our model in three steps. The training process consisted of three stages, as shown in Fig.~\ref{fig:samsung_teacher_student}:
\begin{itemize}
    \item \textbf{Stage 1 – Train Base Model:} Each ERIRNet variant was trained independently using the original ground truth supervision.
    \item \textbf{Stage 2 - Train Teacher Model:} We have trained a Teacher Model based on X-Restomer and fine-tuned the teacher model on each mobile device. 
    \item \textbf{Stage 3 – Distillation with Teacher Model:} We applied knowledge distillation using a teacher model. The model was initialized with the weights trained in the Stage 1. The teacher’s outputs were used as targets to guide both ERIRNet-S and ERIRNet-T.
\end{itemize}

\begin{figure}[ht]
    \centering
    \includegraphics[width=0.98\linewidth]{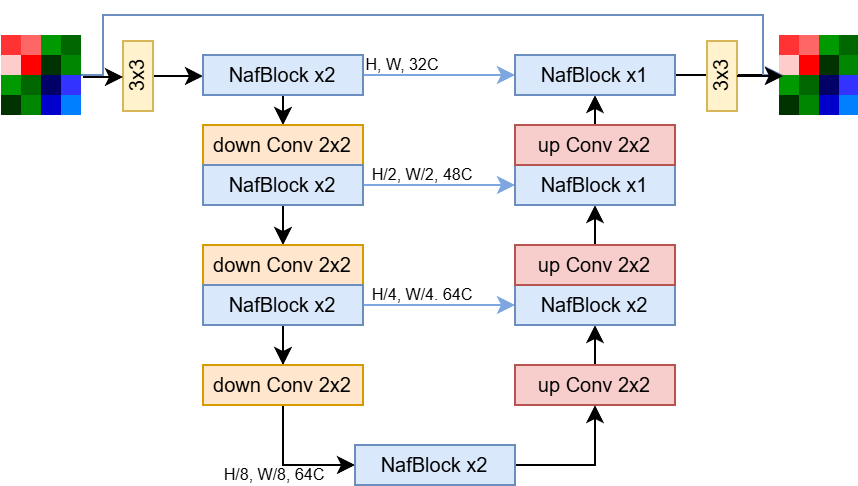}
      \caption{Architectures of ERIRNet-T, with ConvTranspose and reduced FFN expansion. Proposal by Samsung AI.}
    \label{fig:erin_t}
\end{figure}

We used the PyTorch framework for training and utilized A100 GPU. We used the Adam optimizer~\cite{kingma2014adam} with $\beta_1=0.5$, $\beta_2=0.999$ and L1 loss for supervision. The initial learning rate was set to $1e-4$ at stage 1 and reduced to $1e-5$ on stage 3. The model was trained for $1000$, $2000$, and $1000$ epochs in stages 1, 2, and 3, respectively. We implemented a MultiStepLR scheduler to manage learning rate decay, conducting training with a batch size of 16 on NVIDIA A100 GPU.

\subsection{Modified SwinFIR-Tiny for Raw Image Restoration}
\label{sec:xiaomi-ir}

\begin{center}

\vspace{2mm}
\noindent\emph{\textbf{Team Miers}}
\vspace{2mm}

\author{Cheng Li, Lian Liu, Wei Song, Heng Sun \\
Xiaomi Inc. \\
}
\vspace{2mm}
\noindent{\emph{Contact: \url{licheng8@xiaomi.com}}}

\end{center}
\paragraph{Method Description}

Our method is based on an improvement of SwinFIR-Tiny\cite{zhang2022swinfir}. By aggregating the outputs of different RSTB modules to enhance the model's feature representation capability, and also incorporating the HAB module from HAT\cite{chen2023hat} via zero convolution, and applying the reparameterization technique\cite{Gou_2023_ICCV}, the final model has a parameter count of 4.76M and achieved a PSNR of 41.73 on the CodaLab server.

\begin{figure*}[ht]
    \centering
    \includegraphics[width=0.8\textwidth]{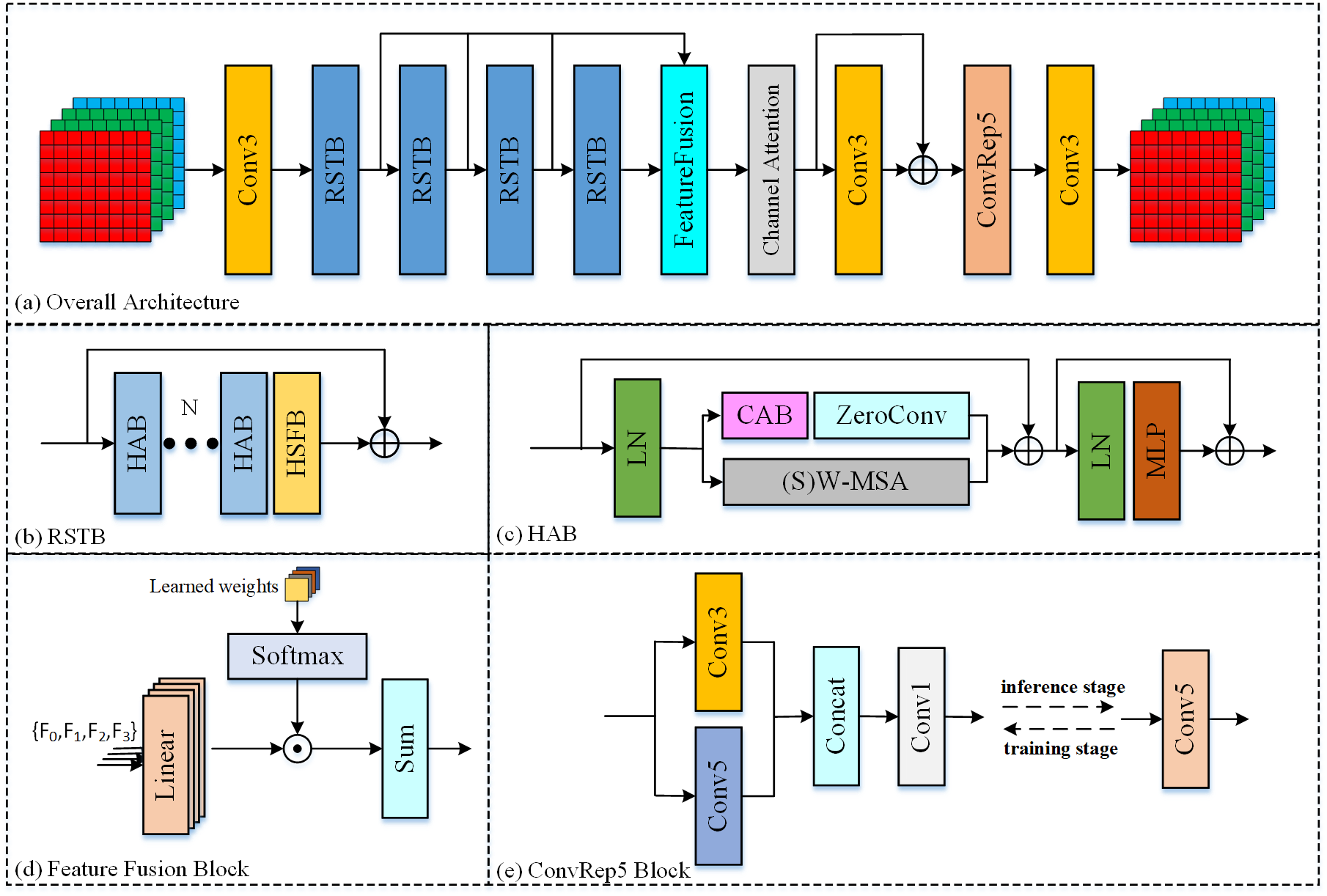}
    \caption{CABATTSwinFIR proposed method by Team Miers (Xiaomi Inc.).}
    \label{fig:mi_model_arch}
\end{figure*}

We selected the SwinFIR-Tiny\cite{zhang2022swinfir} architecture as our baseline framework for evaluating the efficacy of our proposed methodology. The complete network architecture is systematically illustrated in Figure \ref{fig:mi_model_arch}. Our innovative scheme incorporates four Residual Swin Transformer Blocks (RSTBs), which have been extensively validated through empirical analysis to demonstrate exceptional performance in hierarchical feature extraction. Each RSTB consists of 5 or 6 Hybrid Attention Blocks (HABs) and 1 HSFB. Each HAB integrates a multi-head self-attention mechanism and a locally constrained attention module. These complementary mechanisms play a pivotal role in capturing multi-scale contextual information, thereby significantly enhancing the network's representation capabilities across different spatial dimensions.

To further optimize the architectural efficiency, we designed a hierarchical FeatureFusion strategy. This novel structure systematically aggregates the outputs from each RSTB block, effectively mitigating the issue of shallow feature degradation during deep feature extraction. By preserving critical information across hierarchical stages, this fusion mechanism ensures seamless information flow and promotes more discriminative feature abstraction at multiple scales.Experimental evaluations on challenging datasets demonstrate that this architectural configuration achieves substantial performance improvements in handling severely blurred scenes, producing sharper and more visually consistent outputs. 

Complementing these core components, our design integrates specialized enhancements including the Channel Attention Block (CAB\cite{chen2023hat}) and CovRep5\cite{Gou_2023_ICCV} module. These innovations are specifically engineered to improve noise robustness and blur resilience, enabling precise reconstruction of degraded inputs with superior fidelity.

\paragraph{Implementation details}

Our code is implemented using the PyTorch framework and is modified based on the SwinFIR project. The dataset is divided into a training set and a validation set, with 2,099 samples in the training set and 40 samples in the validation set. The data degradation pipeline is sourced from BSRAW\cite{conde2024bsraw}, and we have enhanced the noise levels within it. Specifically, we made the following adjustments: 
\begin{itemize}
    \item The value of \text{log\_max\_shot\_noise} for shot noise was increased from -3 to -2;
    \item For heteroscedastic Gaussian noise, the range of sigma\_1\_range was extended from (5e-3, 5e-2) to (5e-3, 1e-1), and the range of sigma\_2\_range was expanded from (1e-3, 1e-2) to (1e-3, 5e-2).
\end{itemize}

The entire training process was conducted on a machine equipped with four H800 80G GPUs. 
Data augmentation was performed using mixup, and the Adam optimizer was used. The loss function used was the Charbonnier loss. Our approach is based on the SwinFIR-Tiny model and involves four stages of development:

Firstly, we used the original SwinFIR-Tiny model as the baseline, employing the original data degradation method. The initial learning rate was set to 2e-4, with a batch size of 8 and an input size of 180×180. The model was trained for 250K iterations.

Secondly, we added a Feature Fusion module, a Channel Attention module, and a ConvRep5 module to the SwinFIR-Tiny model. The model trained in the first stage was used as the initial parameter set. The original data degradation method was retained, with an initial learning rate of 2e-4, a batch size of 8, and an input size of 180×180. The model was trained for 170K iterations.
Third, building on the second stage, we introduced the CAB module using zero convolution and increased the noise intensity. The initial learning rate was set to 3e-5, with a batch size of 8 and an input size of 180×180. The model was trained for 140K iterations.

Finally, based on the third stage, we further adjusted the initial learning rate to 2e-5, reduced the batch size to 2, and increased the input size to 360×360. The model was trained for 15K iterations. 
The final model that we submitted utilized reparameterization techniques to convert the ConvRep5 module into a standard 5×5 convolution. The total number of parameters in the final model is 4.76M.

\subsection{Multi-PromptIR: Multi-scale Prompt-base Raw Image Restoration }
\label{Multi-PromptIR}

\begin{center}

\vspace{2mm}
\noindent\emph{\textbf{Team WIRTeam}}
\vspace{2mm}

\author{Yubo Wang, Jinghua Wang\\
Harbin Institute of Technology (Shenzhen)\\
}
\vspace{2mm}
\noindent{\emph{Contact: \url{wangyubo@stu.hit.edu.cn}}}

\end{center}

\paragraph{Method Description}

Building on the latest improvements in image restoration research, as presented in \cite{li2024promptcir, chen2024bidirectional}, we introduce the multi-scale prompt-base raw image restoration (Multi-PromptIR) method. It transforms a degraded raw image $\textbf{I} \in \mathbb{R}^{H \times W \times 4}$ into a high-quality, clear image $\bar{\textbf{I}} \in \mathbb{R}^{H \times W \times 4}$. The core of our method is a four layers encoder-decoder architecture, where Transformer Blocks \cite{zamir2022restormer} are integrated to enable cross-channel global feature extraction.
During the encoding stage, drawing from the success of leveraging multi-resolution images in image restoration \cite{chen2024bidirectional}, we incorporate images at reduced resolutions (1/2, 1/4, and 1/8 of the original size). This additional information stream significantly enriches the encoding process. In the decoding phase, we adopt a specialized prompt mechanism \cite{potlapalli2023promptir}, which consists of a Prompt Generation Module (PGM) and a Prompt Interaction Module (PIM).

\begin{figure*}[t]
    \centering
    \includegraphics[width=0.9\linewidth]{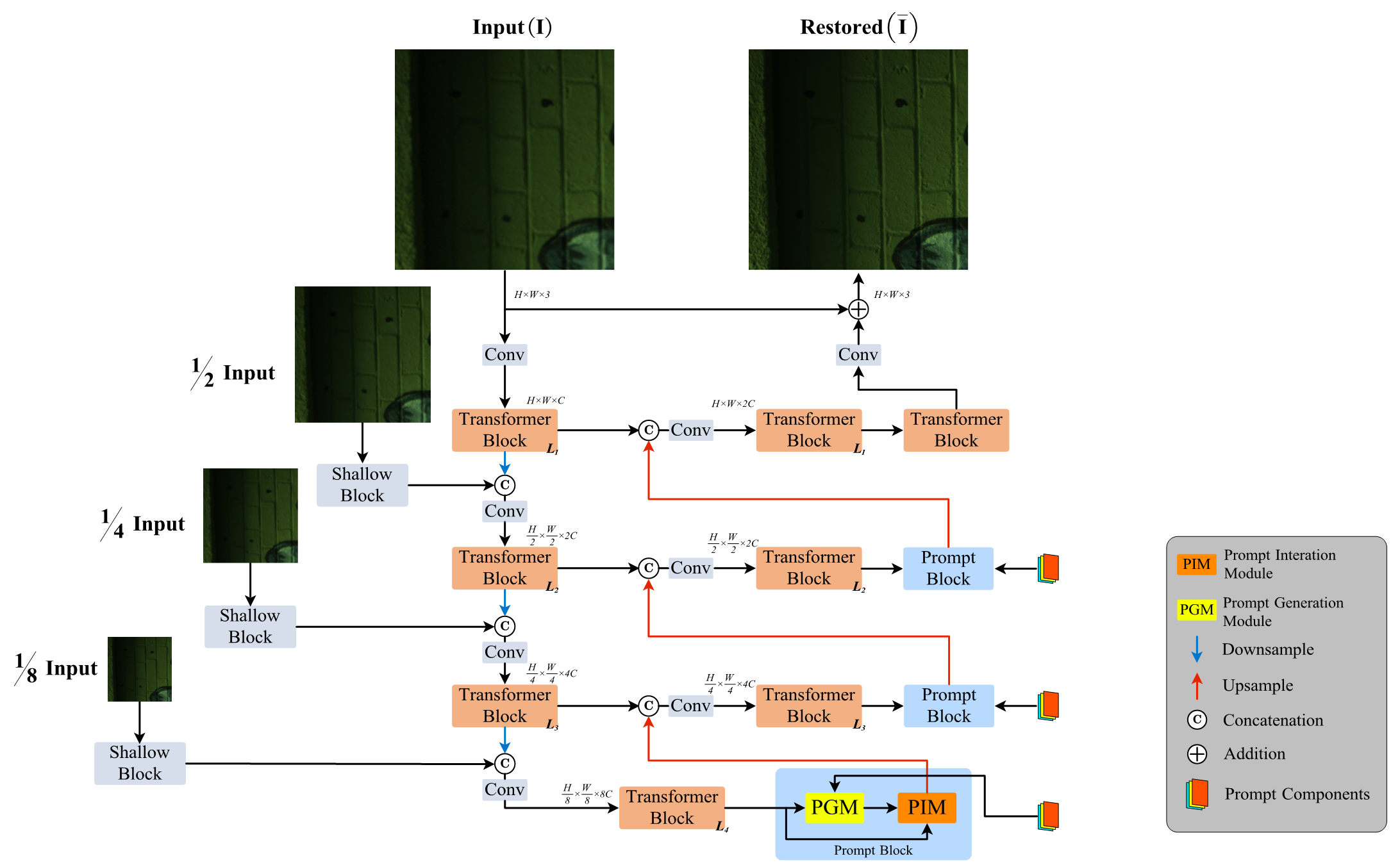}
    \caption{Overall architecture proposed by the Team WIRTeam.}
    \label{fig:Wir_General_model}
\end{figure*}

The model is based PromptIR~\cite{potlapalli2023promptir} and Restormer~\cite{zamir2022restormer}, which combines CNNS and Transformers with specific design demonstrated in Figure \ref{fig:Wir_General_model}. The total number of parameters
in the final model is 39.92 M.

\paragraph{Implementation details}

\begin{itemize}
    
    
    \item \textbf{Optimizer and Learning Rate:} AdamW(the initial learning rate is $2 \times 10^{-4}$ and reduces to $1 \times 10^{-6}$ with the cosine annealing schedule.)
    
    \item \textbf{GPU:} 1 * NVIDIA A100 (80G)
    
    \item \textbf{Datasets:} We merely adopt the datasets provided by organizers. We apply psf blur and add noise to the original images to synthetic degradations.
    \item \textbf{Training and Inference} We train the model in the end-to-end manner for 700 epochs. 
    The training procedure is based on the AdamW optimizer with the decay parameters $\beta_1 = 0.9$ and $\beta_2 = 0.99$.
    The initial learning rate is $2 \times 10^{-4}$ and gradually reduces to $1 \times 10^{-6}$ with the cosine annealing strategy.
    Horizontal and vertical flips are adopted for data augmentation in the dataloader. We split the input degraded images into 256*256 patches and restore them after the process of our proposed architecture. 
        
    \item \textbf{Training Time:} Around 12 hours
    
\end{itemize}

\subsection{LMPR-Net: Lightweight Multi-Stage Progressive RAW Restoration}
\label{LMPR-Net}

\begin{center}

\vspace{2mm}
\noindent\emph{\textbf{Team WIRTeam}}
\vspace{2mm}

\author{Yubo Wang, Jinghua Wang\\
Harbin Institute of Technology (Shenzhen)\\
}
\vspace{2mm}
\noindent{\emph{Contact: \url{wangyubo@stu.hit.edu.cn}}}

\end{center}

\begin{figure*}[ht]
    \centering
    \includegraphics[width=0.8\linewidth]{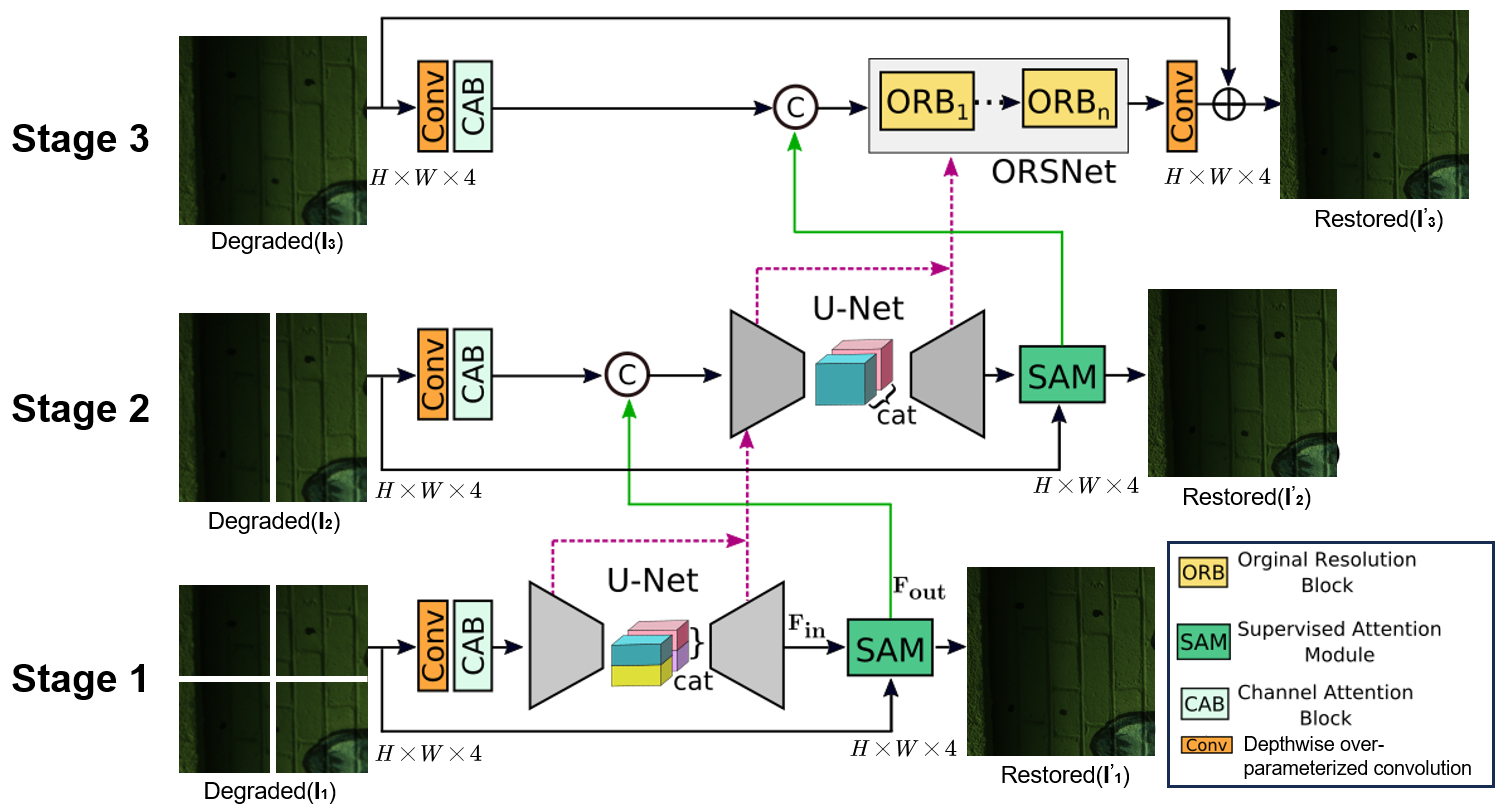}
    \caption{Overall architecture of the LMPR-Net method proposed by the Team WIRTeam.}
    \label{fig:wir_E_framework}
\end{figure*}

\paragraph{Method Description}
Considering the great potential of multi-stage feature interaction in the image restoration process, we propose a lightweight model for raw image restoration based on MPRNet~\cite{zamir2021multi}. The multi-stage model can decompose the raw image restoration task into multiple subtasks to address various degradation information (such as noise, blurring, and other unknown degradations) in this challenge. In the LMPR-Net, the original resolution block (ORB) is composed of convolution and channel attention mechanism to achieve the extraction of cross-channel key features, and the SAM can efficiently refine the incoming features at each stage.
For the sake of lightness, we have simplified some components and set the hidden dimension to be 8. 
Besides, we introduce depthwise over-parameterized convolution~\cite{cao2022conv} to increase the training speed and improve the expressive power of the model without significantly increasing the computational cost.
During the training process, we use the Charbonnier loss~\cite{charbonnier1994two} for constraint and to avoid the restored image from being overly smooth.

We select MPRNet~\cite{zamir2021multi} as the baseline for our method. Our proposed (shown in the Fig~\ref{fig:wir_E_framework} solution builds on MPRNet, introducing targeted improvements that not only enhance performance but also significantly reduce model complexity, thereby delivering a lightweight architecture with low parameter count and flops while keeping excellent raw restoration performance. And the total number of parameters
in the final model is 0.19M with the MACs of 2.63G FLOPs.

\paragraph{Implementation details}
\begin{itemize}
    \item \textbf{Framework:} PyTorch
    \item \textbf{Optimizer and Learning Rate:}  AdamW(the initial learning rate is $2 \times 10^{-4}$ and reduces to $1 \times 10^{-6}$ with the cosine annealing schedule.)
    \item \textbf{GPU:}  NVIDIA GeForce RTX 4090 (24G) * 1
    \item \textbf{Datasets:} We merely adopt the datasets provided by organizers. We apply psf blur and add noise to the original raw images to synthetic degradations. Horizontal and vertical flips are used for data augmentation.
    \item \textbf{Training and Inference} 
    We train our model in an end-to-end manner for 600 epochs. The training process employs the AdamW optimizer with decay parameters set to $\beta_1 = 0.9$ and $\beta_2 = 0.99$. The initial learning rate is set at $2 \times 10^{-4}$, which gradually decreases to $1 \times 10^{-6}$ following a cosine annealing schedule. Data augmentation involves horizontal and vertical flips.
    
    During the inference stage, we partition the input degraded images into $256 \times 256$ patches. After passing through our proposed architecture, these processed patches are seamlessly merged to return to the original size.
    \item \textbf{Training Time:} About 10 hours.
\end{itemize}

\subsection{ER-NAFNet Raw Restoration}
\label{ir_ER-NAFNet}

\begin{center}

\vspace{2mm}
\noindent\emph{\textbf{Team ER-NAFNet}}
\vspace{2mm}

\author{Tianyu Hao~$^1$, Yuhong He~$^2$, Ruoqi Li~$^3$, Yueqi Yang~$^4$ \\ 
$^1$ Huazhong University of Science and Technology \\ 
$^2$ Northeastern University \\
$^3$ Dalian University of Technology \\
$^4$ Institute of Automation, Chinese Academy of Sciences \\

}
\vspace{2mm}
\noindent{\emph{Contact: \url{tianyuhao.times@gmail.com}}}

\end{center}

\paragraph{Method Description}

\begin{figure}[t]
    \centering
    \includegraphics[width=0.98\linewidth]{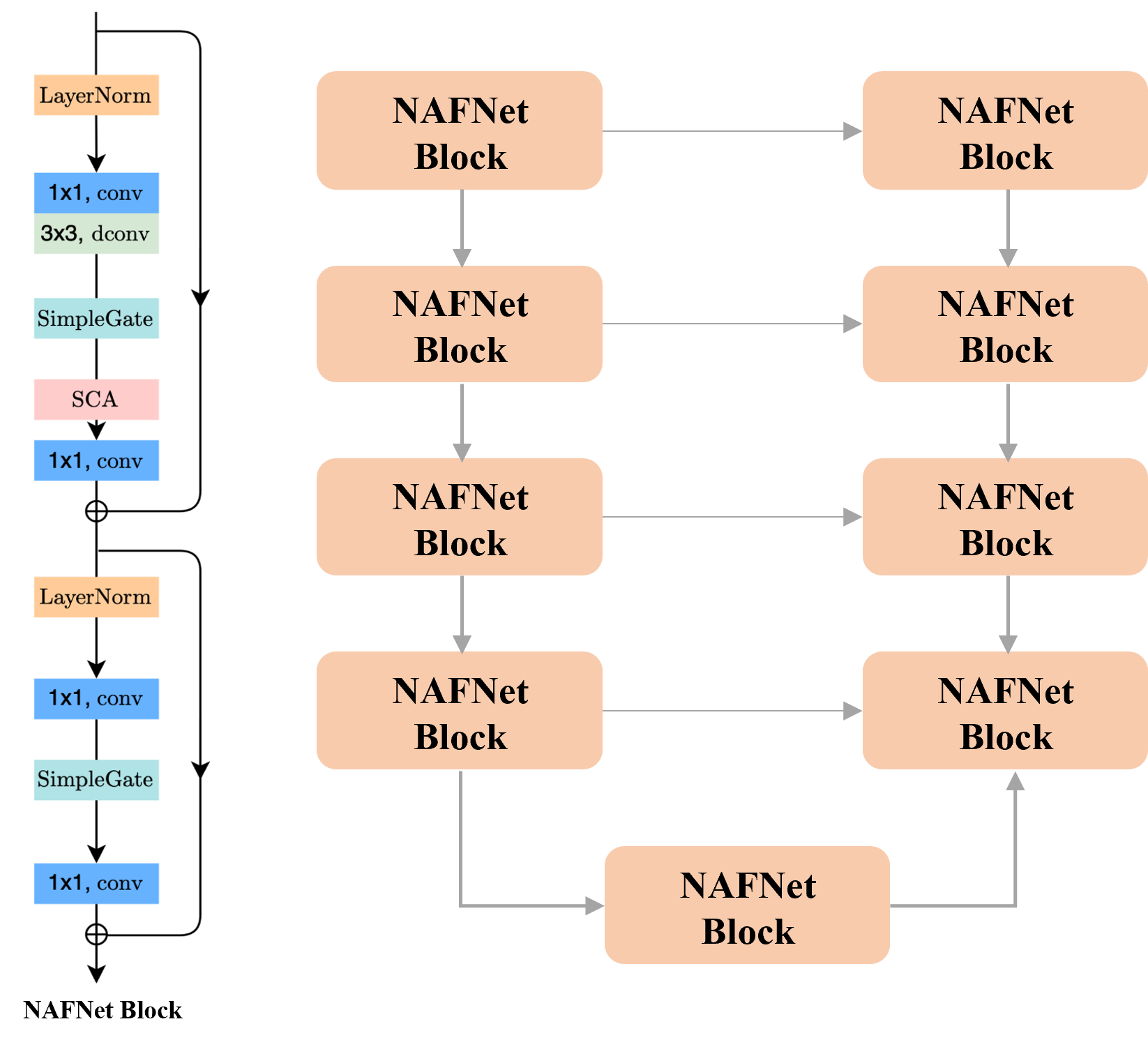}
    \caption{The overall network architecture proposed by Team ER-NAFNet.}
    \label{fig:Team_ChickenRun}
\end{figure}

\begin{figure*}[!ht]
    \centering
    \setlength{\tabcolsep}{1pt} 
    \renewcommand{\arraystretch}{0.5} 
    \begin{tabular}{c c c c c}
         \includegraphics[width=0.19\linewidth]{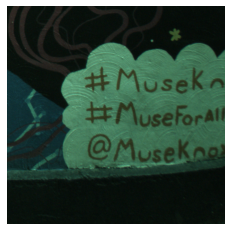} & 
         \includegraphics[width=0.19\linewidth]{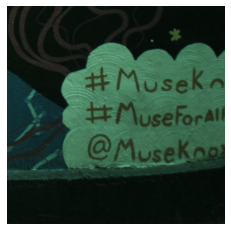} & 
         \includegraphics[width=0.19\linewidth]{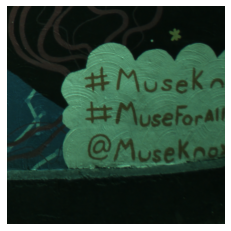} & 
         \includegraphics[width=0.19\linewidth]{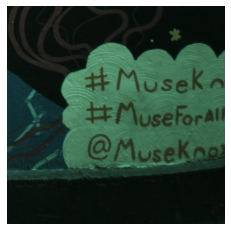} & 
         \includegraphics[width=0.19\linewidth]{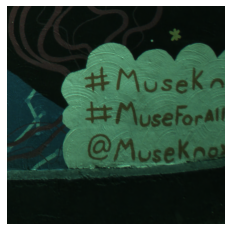}
         \tabularnewline
         Input (Level 1) & PMRID & NAFNET & Samsung AI (E) & WIRTeam (E) \\
         
         \includegraphics[width=0.19\linewidth]{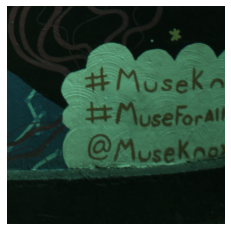} & 
         \includegraphics[width=0.19\linewidth]{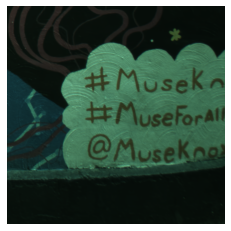} & 
         \includegraphics[width=0.19\linewidth]{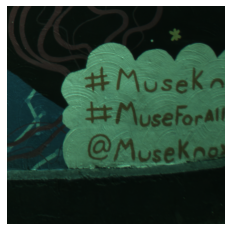} & 
         \includegraphics[width=0.19\linewidth]{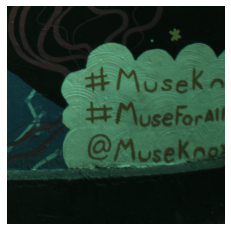} & 
         \includegraphics[width=0.19\linewidth]{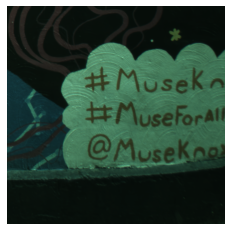}
         \tabularnewline
         MOFA & Samsung AI  & WIRTeam & Team ER-NAFNet & Team Miers \\
         \includegraphics[width=0.19\linewidth]{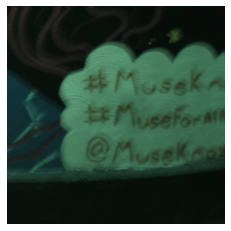} & 
         \includegraphics[width=0.19\linewidth]{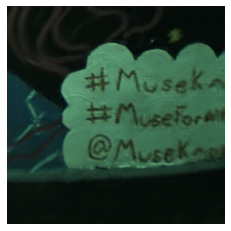} & 
         \includegraphics[width=0.19\linewidth]{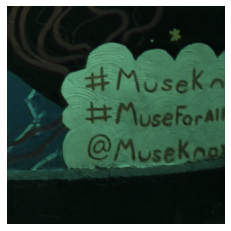} & 
         \includegraphics[width=0.19\linewidth]{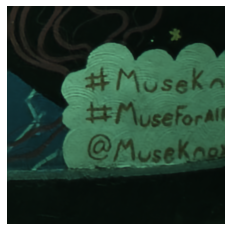} & 
         \includegraphics[width=0.19\linewidth]{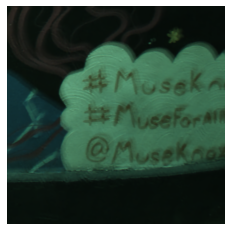}
         \tabularnewline
         Input (Level 2) & PMRID & NAFNET & Samsung AI (E) & WIRTeam (E) \\
         
         \includegraphics[width=0.19\linewidth]{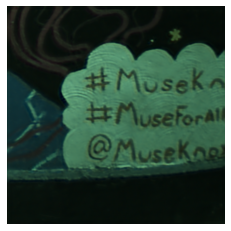} & 
         \includegraphics[width=0.19\linewidth]{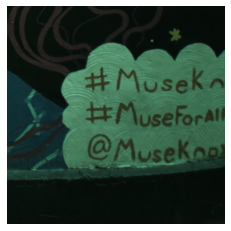} & 
         \includegraphics[width=0.19\linewidth]{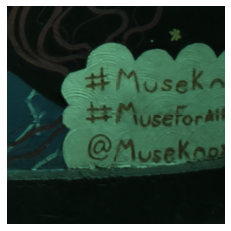} & 
         \includegraphics[width=0.19\linewidth]{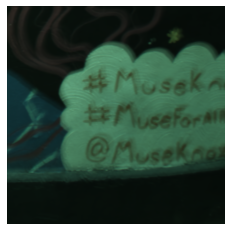} & 
         \includegraphics[width=0.19\linewidth]{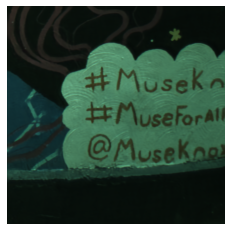}
         \tabularnewline
         MOFA & Samsung AI & WIRTeam & Team ER-NAFNet & Team Miers \\
         \includegraphics[width=0.19\linewidth]{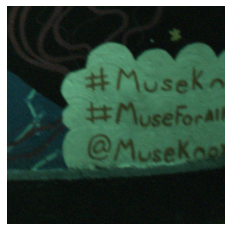} & 
         \includegraphics[width=0.19\linewidth]{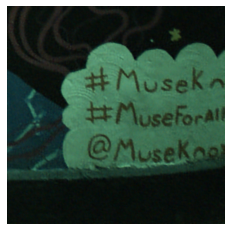} & 
         \includegraphics[width=0.19\linewidth]{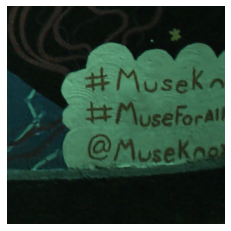} & 
         \includegraphics[width=0.19\linewidth]{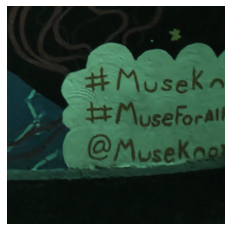} & 
         \includegraphics[width=0.19\linewidth]{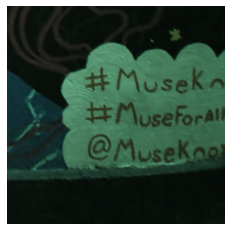}
         \tabularnewline
         Input (Level 3) & PMRID & NAFNET & Samsung AI (E) & WIRTeam (E) \\
         
         \includegraphics[width=0.19\linewidth]{Teams/comp/naf-3.png} & 
         \includegraphics[width=0.19\linewidth]{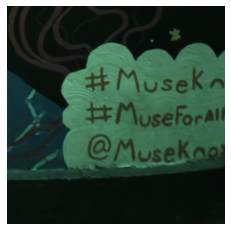} & 
         \includegraphics[width=0.19\linewidth]{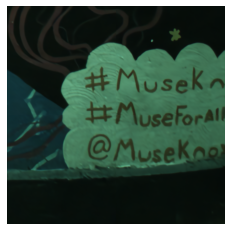} & 
         \includegraphics[width=0.19\linewidth]{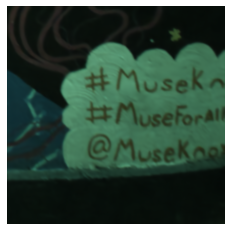} & 
         \includegraphics[width=0.19\linewidth]{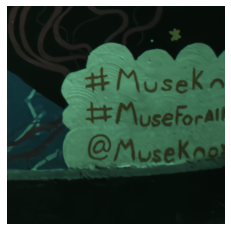}
         \tabularnewline
         MOFA & Samsung AI & WIRTeam & Team ER-NAFNet & Team Miers \\
    \end{tabular}
    \caption{Visual comparison using the \textbf{NTIRE 2025 Challenge on RAW Image Restoration Challenge} testing set (\texttt{gp\_2\_11.npy}). RAW images have $512\times512$ resolution and 4-channels (RGGB Bayer pattern). RAW images are visualized using bilinear demosaicing, gamma correction and tone mapping. (E) indicates efficient methods.}
    \label{fig:results2}
    \end{figure*}

The team introduces ER-NAFNet, a U-shaped  framework designed for Efficient raw image Restoration. The architecture and compression mechanisms of ER-NAFNet are built upon the NAFNet block proposed in NAFNet\cite{NAFNet}(see Fig.~\ref{fig:Team_ChickenRun}), enhancing its efficiency and performance in image restoration task. The ER-NAFNet is trained and learned directly from the 4-channel RGGB RAW data with a complex blurring and noise degradation pipeline, which the noise model, blur kernel and degradation model are demonstrated in AISP\cite{conde2024rawir}. 

Dataset Degradation method, inspired by \cite{ren2020neural} \cite{timofte2015} and \cite{zhang2020deep}, in order to speed up training process and enable the model to learn real degradation, the Bayer pattern RAW images are cropped with 512x512. We use a sequence of degradation operations composed of multiple blurring, hardware specific added noise, random digital gain by luma or exposure compensations. When adding noise, in order to better handle noises in dark aeras, we capture dark frames using multiple mobile sensors to model the real dark current noise. In addition, we remove the pixels in the highlight area in order to prevent highlight color cast.

\paragraph{Model Framework} 
The architectural configuration of the model proposed by Team ChickenRun is illustrated in Fig. \ref{fig:Team_ChickenRun}. This network is structured into three main components: shallow feature extraction, deep feature extraction, and raw reconstruction. Initially, a low-quality raw image $I^{LQ}$ $\in$  $\mathcal{R}^{H \times W\times 4}$ is provided as input. The team applies a 3 $\times$ 3 convolutional filter to extract shallow feature encodings $F^{s}$. Next, a classic U-shaped architecture with skip connections is employed to perform deep feature extraction. Finally, a 3 $\times$ 3 convolution is applied to reconstruct the high-quality (HQ) image.

As depicted in Fig. \ref{fig:Team_ChickenRun}, the NAFNet (Non-Attention Feature Network) Block plays a pivotal role in the U-shaped architecture, addressing the limitations of traditional attention mechanisms in neural networks. The NAFNet Block integrates a series of convolutional operations, including 1x1 and 3x3 dilated convolutions, to effectively capture both fine-grained details and large-scale patterns within the input data. By eliminating the reliance on complex attention mechanisms, the NAFNet Block achieves superior performance with significantly reduced computational overhead. Furthermore, the incorporation of the SimpleGate and Simple Channel Attention (SCA) modules enables the network to focus on the most relevant features while suppressing irrelevant ones, thereby enhancing the overall quality of the feature representations.

\paragraph{Implementation details}

The solution is optimized solely on the NTIRE 2024 official challenge data \cite{conde2024rawir}, using the proposed Development Phase submission set for validation. The experiments are based on the Pytorch framework for implementation. The model uses a width 16, with [2, 2, 4, 8] encoder blocks and [2, 2, 2, 2] decoder blocks at each stage. The middle block num is set to 6.
Since the degraded raw images are characterized by low quality, with a considerable level of details being lost, a data augmentation technique is applied, to improve the training procedure in terms of stability, convergence, and the achieved performance level. The strategy combines simple horizontal or vertical flips with channel shifts and mixup augmentations. The training objective is based on the L2 loss.

The optimization technique used is the AdamW \cite{loshchilov2017decoupled} optimizer ($\beta_{1}$ = 0.9, $\beta_{2}$ = 0.999, weight decay 0.00001) with the cosine annealing strategy, where the learning rate gradually decreases from the initial learning rate $3 \times 10^{-4}$  to $1 \times 10^{-6}$ for 300000 iterations.  The training batch size is set to 12 and patch size is 512.  All experiments are conducted on A100 GPUs.


\clearpage
\newpage
\clearpage

{\small
\bibliographystyle{ieeenat_fullname}
\bibliography{refs}
}

\end{document}